\documentstyle[prb,multicol,epsf,aps]{revtex}
\newcommand{\bi}{\bibitem}
\newcommand{\beq}{\begin{equation}}
\newcommand{\eeq}{\end{equation}}
\newcommand{\bea}{\begin{eqnarray}}
\newcommand{\eea}{\end{eqnarray}}

\newcommand{\ra}{\rangle}
\newcommand{\la}{\langle}
\newcommand{\rperp}{{\bf r}_{\perp}}
\begin{document}
\draft
\title{Gapped phases of quantum wires\footnote{to appear in 
\lq\lq {\it Interactions and Quantum Transport Properties
of Lower Dimensional Systems}\rq\rq\/, Lecture Notes in Physics,
Springer (Proceedings of the 
International WEH Workshop, 
July 1999, Hamburg).}}
\author{Oleg A. Starykh $^a$, Dmitrii L. Maslov $^b$, Wolfgang H\"ausler $^c$,
 and Leonid I. Glazman $^d$}
\address{
$^a$ Department of Applied Physics, Yale University, P. O. Box 208284 \\
New Haven, CT 06520-8284\\
$^b$ Department of Physics, University of Florida, P. O. Box 118440 \\
Gainesville, FL 32611-8440\\
$^c$ Institut f\"ur Theoretische Physik der Universit\"at Hamburg\\
Jungiusstr. 9, D-20355 Hamburg, Germany\\
$^d$ School of Physics and Astronomy\\ University of Minnesota, Theoretical
Physics
Institute\\ 116 Church St., SE Minneapolis, MN 55455}
\maketitle
\centerline{(\today)}
\begin{abstract}
We investigate possible nontrivial phases of a two-subband quantum wire. 
It is found that inter- and
intra-subband interactions may drive the electron system of the wire
into a gapped state. If the nominal electron densities in the two
subbands are sufficiently close to each other, then the leading
instability is the inter-subband 
charge-density wave (CDW).
For large density imbalance, the 
interaction in the inter-subband Cooper channel
may lead to a superconducting instability.
The total
charge-density mode, responsible for the conductance of an ideal
wire, always remains gapless, which enforces the two-terminal
conductance to be at the universal value of $2e^2/h$ per occupied subband.
On the contrary, the tunneling density
of states (DOS) in the bulk of the wire acquires a hard gap, above
which the DOS has a non-universal singularity. This singularity is
weaker than the square-root divergency characteristic for
non-interacting quasiparticles near a gap edge due to the 
\lq\lq dressing\rq\rq\/ of massive modes by a gapless total charge density
mode.
The DOS for tunneling into the end of a
wire in a CDW-gapped state preserves the power-law
behavior due to the frustration the edge introduces into the CDW
order. This work is related to the vast
literature on coupled 1D systems, and most of all, on 
two-leg Hubbard ladders. Whenever possible,
we give derivations of the important results
by other authors, adopted for the context of our study. 

\end{abstract}


\tableofcontents

\section{Introduction}

{From} a theorist's point of view, electrons in quantum wires should provide a
simplest realization of a Luttinger liquid 
\cite{schulz_lezush,fisher_glazman}. Indeed,
as the motion is confined in the direction transverse to the axis of a wire,
the system is effectively one-dimensional; the electron-electron interaction
is strong enough (typically, of the order of the Fermi energy) for the
interaction effects not be washed by the temperature; and the
state-of-the-art wires (at least the semiconductor version of them) are
clean enough for disorder effects to be sufficiently weak. A crucial
experimental test for the existence of the Luttinger-liquid state in quantum
wires would be provided by tunneling into a wire, either in
the middle or into the end. As is well-known, the tunneling density of
states (DOS) of a Luttinger liquid reveals a pseudogap behavior, i.e, it is
suppressed at energies close to the Fermi-energy, which should result
in a power-law bias-dependence of the tunneling current,
and in a power-law temperature dependence of the Ohmic conductance.
Finite-bias and finite-temperature measurements have already been
performed on a very special realization of quantum wires -- carbon
nanotubes, and observed non-linear current-voltage dependences were
interpreted in terms of the Luttinger-liquid theory
\cite{cobden,dekker}. Features of resonant tunneling,
characteristic for a Luttinger liquid, have recently been
observed on GaAs quantum wires \cite{yacoby_new} prepared by cleaved edge
overgrowth technique.
Also, a Luttinger-liquid behavior
has been reported in tunneling into InSb wires naturally
grown in a porous material (asbestos) \cite{zaitsev}. 
Tunneling pseudogap of a
quantum wire has been described in terms of a Luttinger-liquid model both
for a single- \cite{kane_fisher} and multi-subband wire \cite
{matveev_glazman}, the latter system exhibiting a smooth healing of the
pseudogap as the number of the occupied channels increases. In anticipation
of more and better controlled tunneling experiments on quantum wires,
 and also from a general point
of view, we would like to ask if there are any processes which could open a
true gap, rather than a pseudogap, in the electron spectrum of a wire, and if
yes, what are the properties of the corresponding gapped phases.

To this end, we consider in this paper a two-subband quantum wire,
having in mind semiconductor nanostructures studied recently in, e.g.,
Refs.~\cite{tarucha,pepper,yacoby}.
To some approximation, this system is similar to other
two important classes of 1D two-band systems studied extensively
over the last few years, i.e., two-leg Hubbard
ladders \cite{finkelstein,fabrizio,schulz2,leon2,schulz,ekz,balents}
and (single-wall) carbon nanotubes \cite{kane,krotov,egger,gogolin,odintsov}
(for an account of earlier results on coupled 1D systems, see, e.g.,
\cite{prigodin}).
Studies of Hubbard ladders
identified inter-subband scattering processes capable of 
driving the system into a gapped state. Phase diagrams
of a generic ladder, containing a multitude of gapped states, were
constructed in Refs.~\cite{schulz2,leon2,schulz}.
A similarly formulated problem, with applications to a 1D system with
electron and hole bands (valence-fluctuation problem), 
was investigated some time ago in \cite{varma}.

The goal of the present paper is two-fold. First of all, we would like
to understand which of the gapped phases, found in Hubbard ladders,
have a chance to occur in quantum wires. The main difference
between these two systems is that the Coulomb interaction in wires
is (i)~ (supposed to be) purely repulsive; (ii)~relatively long-ranged
(even in the presence of a metallic gate); (iii)~relatively
well-known at distances larger than the lattice spacing
(which is the range relevant for quantum wires); this imposes
constraints on the choice of coupling constants for the Hubbard model.
Also, because the electron wavelength in wires is larger than the
lattice spacing, Umklapp scattering is unimportant.
All these constraints
reduce the variety of possible gapped states
to (a)~inter-subband charge-density wave (CDW), and (b)~superconducting
state. [We will come back to a more detailed description of these
states shortly.] The main difference between carbon nanotubes
and quantum wires is that the former, because of its
special crystal structure, has two conducting subbands
with {\it commensurate} Fermi-momenta.
Although, as we will show,
a quantum wire with nominally different subband Fermi-momenta
may be driven into the commensurate state by
inter-subband backscattering, this process occurs
in a competition with other processes and
requires special analysis. 

Having determined which gapped phase can in principle occur
in a quantum wire, we focus on the calculation of measurable quantities
in each of these phases, which is the second goal of the present
paper. Our main emphasis is on the tunneling density of states,
which turns out to exhibit an unusual threshold
behavior due to coexistence of gapped and gapless modes
and also be sensitive to the presence of open boundaries.
In addition, we consider the two-terminal conductance both
in the absence and in the presence of impurities.

Although this paper is not supposed to be a review, we
present, when possible, derivations of important
results by other authors, e.g., Refs.~\cite{gogolin,voit,wiegmann,orignac},
adopted to the context
of our study. Hopefully, this would help
a reader, who is not an expert in the field,
to understand connections between different
approaches.  

Having formulated the goals and scope of this paper, 
we now return to a generic two-subband
quantum wire with {\it incommensurate} Fermi-momenta  $k_{1F}$ and $k_{2F}$
in subbands 1 and 2, respectively. In the basis
of occupied transverse states, this becomes the problem
of two Luttinger liquids coupled by inter-subband interactions.
To understand possible phases of such a system, one should consider
all possible scattering processes involving electrons from different
subbands, i.e., forward scattering, backscattering, and \lq\lq Cooper
scattering\rq\rq
(cf. Figures \ref{fig:forward},\ref{fig:bs},\ref{fig:cooper}).
Forward scattering simply
renormalizes parameters of Luttinger liquids formed by electrons of each
subbands but does not result in new phases, although it does change the
conditions for occurrence of new phases. (To be more precise, forward
scattering between Luttinger liquids in different subbands is
responsible for the crossover into the Fermi-liquid state, but this
crossover occurs smoothly as the number of channels increases).

The momentum transfer in a backscattering event involving electrons
of different subbands is equal to $k_{1F} \pm k_{2F}$. If
$|k_{1F}-k_{2F}|\gg T/{\rm {min}} \{v_{1F},v_{2F}\}$, then there are
no final states available for electrons involved in such a process
(here $v_{1F}$ and $v_{2F}$ are the Fermi velocities in the two
subbands). Thus, if the temperature $T$ is low enough, interchannel
backscattering is forbidden. However, it may become energetically
favorable for a system to equalize the charge densities, and hence the
Fermi momenta, of different subbands. This may occur if
 the Fermi-momenta difference is
small enough and the amplitude of backscattering is large enough.
After the densities are adjusted, backscattering becomes possible.
As a result, inter-subband charge-density wave (CDW) phase may be
formed, in which charge densities of the subbands form a staggered
pattern, see Fig.~\ref{fig:cdw}.  Similarly to classical
charge-density waves, this phase is very sensitive to a random
potential, resulting in pinning of the CDW and strong suppression of
conductance with disorder.

The \lq\lq Cooper\rq\rq\/ scattering event, on the other hand, always
conserves momentum and energy. In this process, two electrons starting
in, {\it e.g.}, subband 1 with momenta $k_{1F}$ and $-k_{1F}$, scatter on each
other and end up in the other subband, also with opposite momenta $k_{2F}$
and $-k_{2F}$. \lq\lq Cooper scattering\rq\rq\/ can be considered as
formation of a fluctuational Cooper pair in one of the subband
followed by its tunneling into the other one. When kinetic energy
gain due to such tunneling overcomes the Coulomb repulsion,
the wire is in a Cooper (or superconducting)
phase. This phase is characterized by locking of fluctuating charge
currents in different subbands to each other as well as by spin gaps in
each of the subbands.
The Cooper phase is favored when Fermi-momenta imbalance is largest, i.e.
when the second subband just starts to fill up.  Disorder
has a less pronounced effect on Cooper phase than on CDW one, similarly
to what happens in higher dimensions.

It is important to emphasize here that inter-subband backscattering and Cooper
scattering block only modes corresponding to relative charge- and spin-excitations, but leave the center-of-mass charge mode free.
As a result, the conductance remains
at the universal value of $2e^2/h$ per occupied subband in clean CDW and Cooper phases.

Despite being ideal conductors, both the CDW and Cooper phases
are characterized by the truly gapped behavior of the tunneling density
of states at energies
below corresponding gaps.
This is so because a 1D electron is a convolution of charge
and spin collective excitations, and if some of these
excitations acquire a gap, the entire electron 
acquires it
as well.

Somewhat surprisingly, we find that tunneling into the end of the CDW
wire is quite different. A tunnel barrier at the end of the wire distorts
charge-density wave profile and creates a static semi-soliton. This allows 
tunneling into the end to occur even at energies
below the bulk CDW gap (in the lowest order in the barrier's transparency).

In Section \ref{sec:mott} we  consider, for illustrative purposes,
the density of states of the \lq\lq Mott phase\rq\rq\/ , 
which occurs in a single- or multi-subband wire subject to an external 
periodic potential
\cite{mori,starykh_maslov}. In the case of a wire, formed in a semiconductor
heterostructure,
this potential may be provided by an additional electrostatic gate of 
periodic shape \cite{kouwenhoven}. Varying the potential applied to this
gate, one can tune electrons of the wire into the half-filling condition
(one electron per unit cell of the periodic potential). Unlike the two
strong-coupling phases mentioned above, the Mott phase, which is
described by the half-filled Hubbard model, does not conduct
current because its total charge fluctuations are gapped by the external
potential. 

\section{Hamiltonian of a two-subband quantum wire}
\subsection{Classification of scattering processes}
Electrons in a quantum wire are described by the following
Hamiltonian   
\begin{eqnarray}
H &=&\sum_s\int d^{3}r\Psi _{s}^{\dagger }({\bf r})\left(-\frac{1}{2m}{\vec{
\bigtriangledown}_{r}}^{2}-\mu +V_{conf}({\bf r}_{\perp})\right)\Psi _{s}({\bf r})
\nonumber \\
&&+\frac{1}{2}\sum_{s,s'}\int d^{3}rd^{3}r^{\prime }U({\bf r}-{\bf r}^{\prime })
\Psi
_{s}^{\dagger }({\bf r})\Psi _{s^{\prime }}^{\dagger }({\bf r}^{\prime
})\Psi _{s^{\prime }}({\bf r}^{\prime })\Psi _{s}({\bf r}),  \label{ham}
\end{eqnarray}
where $s$ is the spin index, $V_{conf}({\bf r}_{\perp})$ is the confining potential
in the transverse
direction, and $U({\bf r})$ is the electron-electron
interaction potential. The Fermi-wavelength of electrons  is assumed to be 
much larger  than the lattice spacing of the underlying crystal structure.
Because of that, we
do not consider umklapp processes, in which electron momentum is transferred to
the lattice (except for in Ch.\ref{sec:mott}). 
Hamiltonian (\ref{ham}) is Galilean-invariant, and hence our
subsequent calculations have to preserve this invariance as well. We will
return to this important point later on in our discussion.

If the chemical potential in the leads is such that only two lowest
 subbands of transverse quantization are occupied,
the electron wavefunction is given by 
\begin{equation}
\Psi _{s}({\bf r})=\sum_{n=1}^{2}\phi_{n}(\rperp)\psi _{ns}(x),
\label{trans-wf}
\end{equation}
where $\phi_{n}(\rperp)$ are the orthogonal wavefunctions of
transverse quantization, chosen to be real.
 In this basis, the
kinetic part of Hamiltonian (\ref{ham}) becomes 
\begin{equation}
H_{0}=\sum_{n,s}\int dx\psi _{ns}^{\dagger }(x)\left(-\frac{\partial _{x}^{2}}{2m}
-\mu +\epsilon _{ns}\right)\psi _{n}(x)
\end{equation}
where $\epsilon _{n}$
is the energy of the n-th transverse subband.

To describe low-energy excitations in the n-th subband, we expand
the longitudinal part of the $\Psi$-operator, $\psi_s(x)$,
in terms of right- and left-moving excitations,
residing around the Fermi-points of the n-th channel:
\begin{equation}
\psi _{ns}(x)=R_{ns}(x)e^{ik_{nF}x}+L_{ns}(x)e^{-ik_{nF}x}.  \label{bos}
\end{equation}

\begin{figure}[t]
\hspace*{5cm}
\epsfxsize=5cm
\epsfbox{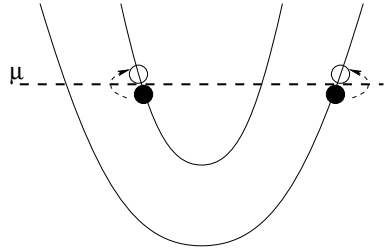}
\vskip0.5cm
\caption{Example of inter-subband forward scattering.
Filled (empty) circles  denote initial (final) states of electrons.
Dashed lines with arrows indicate \lq \lq direction\rq\rq of the scattering.
}
\label{fig:forward}
\end{figure}

In this representation, the interaction  (four-fermion) part of 
Hamiltonian (\ref{ham}) reduces to a sum of two terms.
The first one, $U_{intra}$, describes
the interaction of electrons within the same subband, and contains
forward and backward scattering processes. The second one, $U_{inter}$, describes
the inter-subband interaction. It splits naturally into ${\it forward}$ ($U^{F}$)%
${\it ,}$ {\it backward (}$U^{B}$){\it , }and {\it Cooper (}$U^{C}$) parts.

\begin{equation}
U_{inter}=U^{F}+U^{B}+U^{C}.
\end{equation}
The momentum transfer between subbands is limited by the width of the Fermi distribution function, and is therefor small at low temperatures. Forward scattering involves no momentum transfer between subbands
(cf. Fig.\ref{fig:forward}). This process is also an example of a 
{\em direct} process,
in a sense that electrons stay in the same subband, as is evident
from the explicit expression for $U^{F}$ 
\begin{eqnarray}
U^{F} &=&\frac{1}{2}\sum_{n\neq m}\int_{x,x^{\prime
}}M_{d}^{\{nm\}}(x-x^{\prime })\sum_{s,s^{\prime }}[R_{ns}^{\dagger
}(x)R_{ns}(x)+L_{ns}^{\dagger }(x)L_{ns}(x)]  \nonumber
\\
&&\times \lbrack R_{ms^{\prime }}^{\dagger }(x^{\prime })R_{ms^{\prime
}}(x^{\prime })+L_{ms^{\prime }}^{\dagger }(x^{\prime })L_{ms^{\prime }}
(x^{\prime })],
\end{eqnarray}
where the {\it direct} matrix element is given by

\begin{equation}
M_{d}^{\{nm\}}(x-x^{\prime })
=\int_{\rperp,\rperp^{\prime }}U({\bf r}-{\bf r}^{\prime
})\phi _{n}^{2}(\rperp)\phi _{m}^{2}(\rperp');  \label{mdir}
\end{equation}
and $\int_{z,z^{\prime }}\equiv \int dz\int dz^{\prime }$.

\begin{figure}[t]
\hspace*{3cm}
\epsfxsize=10cm
\epsfbox{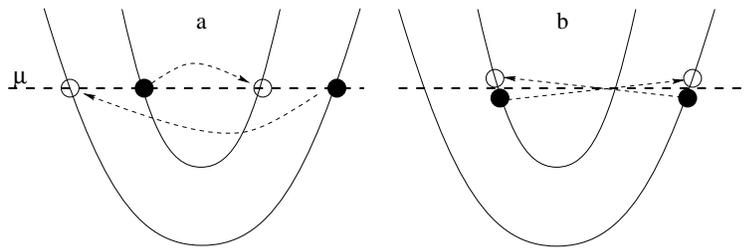}
\vskip0.5cm
\caption{Example of inter-subband backscattering: (a) direct, (b) exchange. 
Notations as in Fig.\ref{fig:forward}.}
\label{fig:bs}
\end{figure}

\begin{figure}[t]
\hspace*{3cm}
\epsfxsize=10cm
\epsfbox{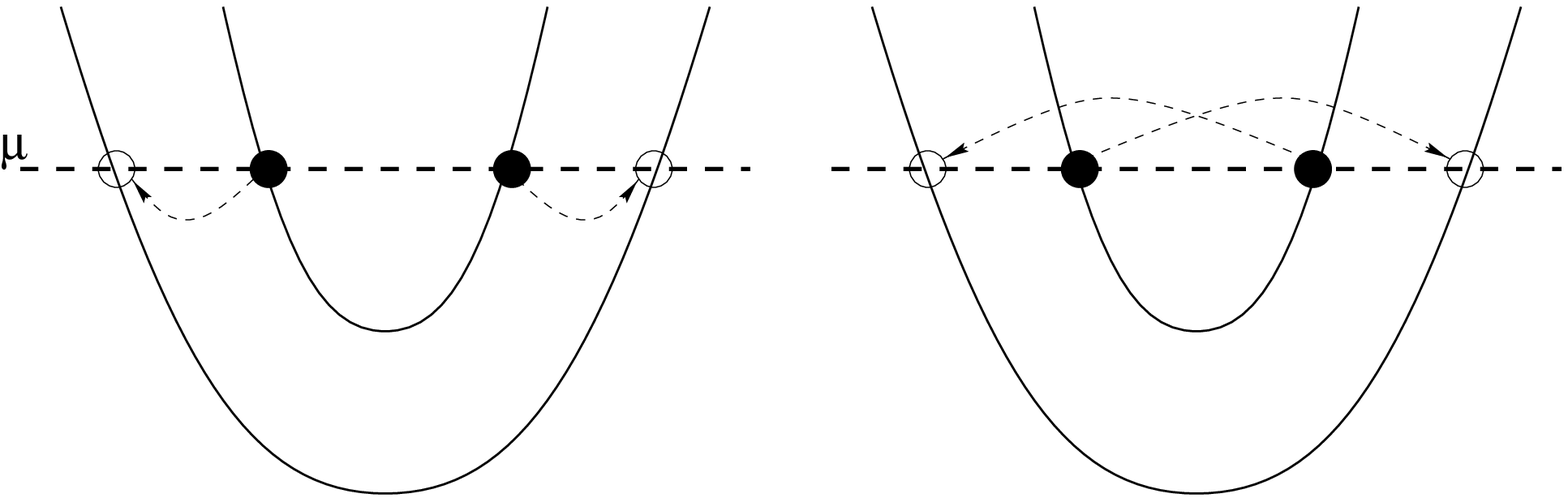}
\vskip0.5cm
\caption{Examples of inter-subband Cooper scattering.
Notations as in Fig.\ref{fig:forward}.}
\label{fig:cooper}
\end{figure}

 By \lq\lq backward
scattering\rq\rq, we understand processes with a non-zero momentum transfer $
\delta k=k_{1F}\pm k_{2F}$ between subbands. These processes can be
divided further into {\it direct }and {\it exchange }parts

\begin{equation}
U^{B}=U_{d}^{B}+U_{x}^{B}.
\end{equation}
In an {\it exchange } process [see Figs.\ref{fig:bs},\ref{fig:cooper}], 
electrons change subbands. Two
parts of backscattering can be written as (for the sake of brevity, we omit
here the $x-x^{\prime }$-dependence of the matrix elements):

\begin{eqnarray}
U_{d}^{B} &=&\frac{1}{2}\sum_{n\neq m}\sum_{s,s^{\prime }}\int_{x,x^{\prime
}}M_{d}^{\{nm\}}[R_{ns}^{\dagger }(x)L_{ns}(x)L_{ms^{\prime }}^{\dagger
}(x^{\prime })R_{ms^{\prime }}(x^{\prime })e^{2i(k_{mF}-k_{nF})(x+x^{\prime
})}  \nonumber \\
&&+(R\Leftrightarrow L)e^{-2i(k_{mF}-k_{nF})(x+x^{\prime })}]
\label{backdir}
\end{eqnarray}
and 
\begin{eqnarray}
U_{x}^{B} &=&-\frac{1}{2}\sum_{n\neq m}\sum_{s,s^{\prime }}\int_{x,x^{\prime
}}M_{x}^{\{nm\}}[\{R_{ns}^{\dagger }(x)R_{ns^{\prime }}(x^{\prime
})R_{ms^{\prime }}^{\dagger }(x^{\prime
})R_{ms}(x)e^{i(k_{mF}-k_{nF})(x-x^{\prime })}  \nonumber \\
&&+(R\Leftrightarrow L)e^{-i(k_{mF}-k_{nF})(x-x^{\prime })}\}  \nonumber \\
&&+\{R_{ns}^{\dagger }(x)R_{ns^{\prime }}(x^{\prime })L_{ms^{\prime
}}^{\dagger }(x^{\prime })L_{ms}(x)e^{i(k_{mF}+k_{nF})(x-x^{\prime
})}+(R\Leftrightarrow L)e^{-i(k_{mF}+k_{nF})(x-x^{\prime })}\}],
\label{backex}
\end{eqnarray}

where the exchange matrix element is 
\begin{equation}
M_{x}^{\{nm\}}(x-x^{\prime })=\int_{\rperp,\rperp^{\prime }}
U(|{\bf r}-{\bf r}^{\prime}|)
\phi _{n}(\rperp)\phi _{m}(\rperp^{\prime })\phi_{n}(\rperp^{\prime })
\phi _{m}(\rperp).
\label{mex}
\end{equation}

Momentum conservation requires that the energy of at least one of the states, 
involved into direct backscattering, should be far away from the Fermi energy, 
which forbids
this process at not too high temperatures 
($T~\ll ~|k_{mF}-k_{nF}|$~min~$\{v_{nF},v_{mF}\}$).
This is reflected in the presence of the exponential factors in front of the
fermion operators in Eq.(\ref{backdir}), which oscillate rapidly
as functions of $(x+x^{\prime })$. This restriction can be lifted though,
if the system prefers to gain energy from backscattering by equalizing the
subband densities, so that $k_{mF}=k_{nF}$.

Finally, we call \lq\lq Cooper scattering\rq\rq (Fig.\ref{fig:cooper}) 
a process in which two
electrons with zero total momentum\ (a fluctuational Cooper pair) hop from,
e.g., channel $m$ , into channel $n$, so that the total momentum $
Q=-k_{m}+k_{m}=0\rightarrow -k_{n}+k_{n}=0$ is conserved. This process is
also  referred to as \lq \lq Josephson coupling\rq \rq 
~\cite{efetov,prigodin}, 
or as \lq\lq $g_{00\pi \pi }$ process\rq\rq
 ~\cite{fabrizio,schulz}.
The Hamiltonian of Cooper scattering is given by
\begin{eqnarray}
U_{{}}^{C} &=&\frac{1}{2}\sum_{n\neq m}\sum_{s,s^{\prime }}\int_{x,x^{\prime
}}M_{x}^{\{nm\}}[R_{ns}^{\dagger }(x)L_{ns^{\prime }}^{\dagger }(x^{\prime
})e^{ik_{nF}(x^{\prime }-x)}+L_{ns}^{\dagger }(x)R_{ns^{\prime }}^{\dagger
}(x^{\prime })e^{-ik_{nF}(x^{\prime }-x)}]  \nonumber \\
&&\times \lbrack R_{ms^{\prime }}(x^{\prime })L_{ms}(x)e^{ik_{mF}(x^{\prime
}-x)}+L_{ms^{\prime }}(x^{\prime })R_{ms}(x)e^{-ik_{mF}(x^{\prime }-x)}].
\label{cooper}
\end{eqnarray}
By construction, Cooper scattering is of the {\em exchange} type.

In what follows, we  use the following abbreviations:
 forward scattering $\equiv $ FS, direct backward scattering $
\equiv $ dBS, exchange backward scattering $\equiv $ xBS, Cooper scattering $
\equiv $ CS.

\begin{figure}[t]
\hspace*{3cm}
\epsfxsize=8cm
\epsfbox{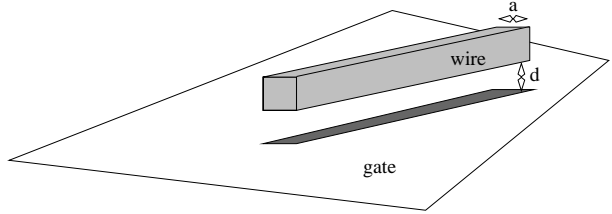}
\vskip0.5cm
\caption{Schematic view of a gated wire. The wire of 
typical transverse size $a$ is separated by distance $d$ from the metallic
gate. Distance $d$ determines the range of interaction among electrons
inside the wire.}
\label{fig:wire}
\end{figure}

For a generic situation of $k_{n}\neq k_{m}$, the only
momentum-conserving inter-subband scattering processes are FS, xBS, and CS.
The amplitudes of these processes
depend on the ratio $a/d$ , where $a$ is a typical transverse
size of the wire (which determines the spatial extent of
$\phi_{n}(\rperp)$) and  $d$ is the interaction range. 
In the limit of $a/d\rightarrow 0$, the interaction
potential can be taken out of integrals (\ref{mdir}) and (\ref{mex}), upon
which $M_{d}$ remains finite, whereas $M_{x}$ vanishes. It can be readily
shown that for finite but small ratio $a/d$, the exchange matrix element is small:
$M_{x}\sim
(a/d)^{2}M_{d}$. The long-range interaction thus discriminates against exchange processes.
If (as it is most often the case) a wire is formed by means of a gate
deposited over the 2D heterostructure, $d$ is given by the distance to the
 gate, which screens the Coulomb interaction in the wire (see Fig.~\ref{fig:wire}). 
Typically,
$a/d=0.1-1$. 

\subsection{Bosonized form of the Hamiltonian}
We use the conventional bosonization procedure,  in which
\begin{eqnarray}
R_{ns}(x) &=&\frac{1}{\sqrt{2\pi \alpha }}e^{i\sqrt{\pi }(\varphi
_{ns}-\theta _{ns})}, \label{bosr} \\
L_{ns}(x) &=&\frac{1}{\sqrt{2\pi \alpha }}e^{-i\sqrt{\pi }(\varphi
_{ns}+\theta _{ns})},  \label{bosl}
\end{eqnarray}
and short-range cut-off $\alpha \sim k_{F}^{-1}$. Boson fields $\varphi _{ns}$ and
$\theta _{ns}$ with
$n=1,2;~s=\pm 1$, are decomposed into charge-($\rho $) and
spin-($\sigma $) 
collective modes 
\begin{eqnarray}
&&\varphi _{ns}=\frac{1}{\sqrt{2}}(\varphi _{n\rho }+s\varphi _{n\sigma }), 
\nonumber \\
&&\theta _{ns}=\frac{1}{\sqrt{2}}(\theta _{n\rho }+s\theta _{n\sigma }).
\label{decomp}
\end{eqnarray}
Parts of the Hamiltonian, corresponding to free motion and intra-subband interactions ($H_0+U_{intra}$), take
the well-known Luttinger-liquid form:
\begin{eqnarray}
H_{n\rho } &=&\frac{1}{2}\int dx\{v_{n\rho }K_{n\rho }(\partial _{x}\theta
_{n\rho })^{2}+\frac{v_{n\rho }}{K_{n\rho }}(\partial _{x}\varphi _{n\rho
})^{2}\},  \label{charge-density} \\
H_{n\sigma } &=&\frac{1}{2}\int dx\{v_{n\sigma }K_{n\sigma }((\partial
_{x}\theta _{n\sigma })^{2}+\frac{v_{n\sigma }}{K_{n\sigma }}(\partial
_{x}\varphi _{n\sigma })^{2}\}  \nonumber \\
&&+\frac{2U(2k_{nF})}{(2\pi \alpha )^{2}}\int dx\cos [\sqrt{8\pi }\varphi
_{n\sigma }],  \label{spin-density}
\end{eqnarray}
which describes {\it independent } charge- and
spin-density excitations ($H_{n\rho }$ and  $H_{n\sigma }$,
respectively).
The cosine term in Eq.(\ref{spin-density}) is due to backscattering within a
single subband. Explicit expressions for the Luttinger-liquid parameters will be
discussed later.

Upon bosonization, the three types of intersubband interactions take the following form~:
\begin{equation}
U^{F}=\frac{2f_{0}}{\pi }\int dx\partial _{x}\varphi _{1\rho
}\partial _{x}\varphi _{2\rho };  \label{forward}
\end{equation}
\begin{eqnarray}
U_{d}^{B} &=&\frac{4f_{bs}}{\pi ^{2}\alpha ^{2}}\int dx\cos \left[ \sqrt{%
2\pi }\left( \varphi _{1\rho }-\varphi _{2\rho }\right) +2(k_{1F}-k_{2F})x%
\right]  \nonumber \\
&&\times \cos [\sqrt{2\pi }\varphi _{1\sigma }]\cos [\sqrt{2\pi }\varphi
_{2\sigma }];  \label{dir_backward}
\end{eqnarray}
\begin{eqnarray}
U_{x}^{B} &=&-\frac{1}{2}\int dx\Big(\frac{b_{1}+b_{2}}{\pi }(\partial
_{x}\varphi _{1\rho }\partial _{x}\varphi _{2\rho }+\partial _{x}\varphi
_{1\sigma }\partial _{x}\varphi _{2\sigma })+\frac{b_{1}-b_{2}}{\pi }%
(\partial _{x}\theta _{1\rho }\partial _{x}\theta _{2\rho }+\partial
_{x}\theta _{1\sigma }\partial _{x}\theta _{2\sigma })  \nonumber \\
&&+\frac{2}{\pi ^{2}\alpha ^{2}}\cos [\sqrt{2\pi }(\theta _{1\sigma }-\theta
_{2\sigma })]\{(b_{1}+b_{2})\cos [\sqrt{2\pi }\varphi _{1\sigma }]\cos [%
\sqrt{2\pi }\varphi _{2\sigma }]-  \nonumber \\
&&(b_{1}-b_{2})\sin [\sqrt{2\pi }\varphi _{1\sigma }]\sin [\sqrt{2\pi }%
\varphi _{2\sigma }]\}\Big)  \label{backward};
\end{eqnarray}
\begin{eqnarray}
U^{C} &=&\frac{4}{2\pi ^{2}\alpha ^{2}}\int dx\{t_{sp}\cos [\sqrt{2\pi }%
(\theta _{1\rho }-\theta _{2\rho })]\cos [\sqrt{2\pi }\varphi _{1\sigma
}]\cos [\sqrt{2\pi }\varphi _{2\sigma }]+  \nonumber \\
&&+t_{tp}\cos [\sqrt{2\pi }(\theta _{1\rho }-\theta _{2\rho })]\Big(\cos [%
\sqrt{2\pi }(\theta _{1\sigma }-\theta _{2\sigma })] - \sin [\sqrt{2\pi }%
\varphi _{1\sigma }]\sin [\sqrt{2\pi }\varphi _{2\sigma }]\Big)\}.
\label{singlet-triplet}
\end{eqnarray}
The corresponding amplitudes are given by
\begin{eqnarray}
f_{0} &=&\int dxM_{d}^{\{12\}}(x),  \nonumber \\
f_{bs} &=&\int dxM_{d}^{\{12\}}(x)\cos [(k_{1F}+k_{2F})x],  \nonumber \\
b_{1,2} &=&\int dxM_{x}^{\{12\}}(x)\cos [(k_{1F}\mp k_{2F})x],  \nonumber \\
t_{sp} &=&\int dxM_{x}^{\{12\}}(x)\cos (k_{1F}x)\cos (k_{2F}x),  \nonumber \\
t_{tp} &=&\int dxM_{x}^{\{12\}}(x)\sin (k_{1F}x)\sin (k_{2F}x).
\label{matrixelements}
\end{eqnarray}
In the last two lines, $t_{sp}(t_{tp})$ are the amplitudes of singlet 
(triplet) Cooper processes.

The highly non-linear (cosine) terms in Eqs.(\ref{dir_backward},\ref
{backward},\ref{singlet-triplet}) signal potential instabilities of the
ground state due to inter-subband interactions. For the dBS process
 [Eq.(\ref{dir_backward}%
)], this instability is of the charge-density-wave (CDW) type, 
quantity $\varphi
_{1\rho }-\varphi _{2\rho }$ being the phase of the CDW (particle-hole)
condensate. If subbands are equivalent ($k_{1F}= k_{2F}$), the energy is
minimized by adjusting the CDW-condensate phase is such a way that the
cosine takes its minimum value ($-1$, for repulsive interactions). For
non-equivalent subbands ($k_{1F}\neq k_{2F}$), the global minimization of
the energy is impossible due to the position-dependent phase shift, 
 and thus the CDW instability is suppressed.
Nevertheless, if the energy gain due to opening of the CDW gap is large
enough, the system may choose to adjust the subband densities, which makes
the CDW instability possible. Density adjustment is most likely to occur
if the cross-section of the wire is approximately symmetric. 
For example, if it is a perfect square, the second subband of transverse
quantization is doubly
degenerate. Deviations from the ideal shape lift the degeneracy
but the energy splitting between the states remains small for small
deviations. Such states are almost equally occupied,
and a small difference in densities is likely to be
eliminated by opening of the CDW-gap. It seems that cleaved edge quantum
wires investigated by Yacoby et al.\cite{yacoby} satisfy this requirement.

In the context of two-leg Hubbard ladders, the CDW-process of this type is known
as \lq\lq deconfinement\rq\rq \cite{fabrizio}: degeneracy of bonding and 
antibonding
subbands implies that the amplitude of interchain tunneling, $t_{\perp }$, is 
renormalized to zero by interactions, and electrons thus remain 
\lq\lq confined\rq\rq
to their respective chains.

The xBS process (\ref{backward}) contains both harmonic terms, arising from
backscattering of electrons with parallel spins, and cosine terms, arising
from backscattering with antiparallel spins. The latter contain
only spin fields and thus can lead to the instability only in the spin
channel. In the terminology of Ref. \cite{schulz,schulz2}, this instability
corresponds to the \lq\lq orbital antiferromagnet phase\rq\rq (OAF).
The OAF instability occurs only if backscattering is sufficiently strong 
\cite{schulz,schulz2}.
For a quantum wire, in which all amplitudes are given just by
the corresponding Fourier components of the same interaction potential, 
this conditions
means that $U(2k_F)>2U(0)$ (for identical subbands), which is never the case 
for any physical $U({\bf r})$. In what follows, we will not
therefore consider the OAF phase.
 (Note that for a Hubbard ladder, amplitudes 
of various
scattering processes may be determined by entirely different physics, e.g.,
some of them may result from direct electron-electron interaction and some 
from
exchange of virtual phonons. Hence, the ratios of amplitudes may be arbitrary,
 and the OAF
instability is possible, at least {\it a priori}.)
  
Finally, the CS process [Eq.(\ref{singlet-triplet})] may lead to a 
superconducting
instability (of both singlet and triplet types), accompanied by opening
of spin gaps in each of the subbands, in a analogy with a superconducting
transition in higher dimensions. Quantity $\theta _{1\rho }-\theta _{2\rho }$
plays the role of the superconducting condensate phase. Inter-subband forward 
scattering [Eq.~(\ref{forward})] has an important role in developing a 
superconducting
instability--it reduces electron repulsion
in the relative charge-density fluctuation channel, making it possible for
Cooper scattering to become relevant.
The superconducting phase is also known
as  a \lq\lq C1S0-phase\rq\rq\/\cite{leon2} (meaning: one gapless charge mode and no 
gapless spin modes)
 or 
a \lq\lq d-wave superconductor\rq\rq\/ \cite{schulz2,schulz}(indicating that the order parameter is odd 
upon interchanging the electrons, forming a Cooper pair, between subband) 
.
This particular instability received much attention recently as one 
of the models of HTC superconductivity \cite{ekz}.
We also note in passing
that the idea of superconductivity in a two-band system has a long history,
starting from the 1968 paper by Fr\"olich \cite{froelich} (for a review, see
Ref.~\cite{ruvalds}). The idea, employed in earlier work, is that
if the masses of electrons in two subbands are significantly
different, there always--even in 3D--exists a gapless plasmon excitation
(a direct analog of Langmuir-Tonks ion sound waves in plasma),
which serves as a mediator of effective attraction.
The superconducting phase in a 1D two-band system is
already composed of gapless excitations and is not
limited by the condition of different masses (although,
as we will see shortly, there is no lack of other constraints).

Note also that $U_{d}^{B}$ (\ref{dir_backward}) and $U^{C}$ (\ref
{singlet-triplet}) mix charge and spin modes, and thus spoil the
spin-charge separation present in the Hamiltonian of a single subband.

Processes (\ref{forward} - \ref{singlet-triplet})
have been written down in the literature in many different ways, so it
is worth to make a connection to previous work here.
Identification of our notations with the \lq\lq $g$-ological\rq\rq\/ ones (used by Schulz in his
two important papers \cite{schulz2,schulz}) is as follows: 
$t_{sp}=g_{12} + g_{23}$,
$t_{tp}=g_{23}-g_{12}$, $b_1=g_{13}$, and his $g_{11}$-process 
(backscattering with opposite spins)
should be equated with our $U(2k_{nF})$ in Eq.(\ref{spin-density}).
There is no correspondence to our amplitude $b_2$, which describes
exchange inter-subband backscattering of the type 
$R^{\dagger}_{ns}R^{\dagger}_{ms'}
R_{ns'}R_{ms} + (R \rightarrow L)$, see Eq.(\ref{backex}). That such a process
is absent in Ref.~\cite{schulz} is clear from Eq.(2) of this reference.

One more connection is made by noting that 
(somewhat lengthy) Eq.(\ref{backward}) can be
represented compactly as 
\begin{equation}
U_{x}^{B}=-\int dx\{(b_{1}+b_{2})(\rho _{1}\rho
_{2}+S_{1}S_{2})+(b_{1}-b_{2})(j_{c1}j_{c2}+j_{s1}j_{s2})\},
\end{equation}
where $\rho_n ~(S_n)$ is charge (spin) density, and $j_{cn}~(j_{sn})$ is charge
(spin) current in the $n$-th subband, using notations of Emery,
Kivelson, and  Zachar \cite{ekz}.

\section{Spinless electrons}
\label{spinless}

\subsubsection{Model}

In this section we consider a \lq \lq toy\rq\rq model of a two-subband 
system of spinless
electrons, which contains all interesting effects we want to discuss
and, at the same time, allows for a rather complete analytic treatment.
In this model, various parts of the Hamiltonian reduce to 
\beq
H_0\to {\tilde H}_0=\int dx\sum_{n}\left[\frac{v_{n}}{2K_{n}}
(\partial _{x}\varphi _{n})^{2}
+\frac{v_{n}K_{n}}{2}(\partial _{x}\theta
_{n})^{2}\right];
\label{h0spinless}\eeq
\beq
U^F\to{\tilde U}^F=\frac{f_{0}}{
\pi }\int dx\partial _{x}\varphi _{1}\partial _{x}\varphi _{2};
\label{f-spinless}
\eeq
\begin{equation}
U_d^B\to{\tilde U}_{d}^{B}=
\frac{f_{bs}}{\pi ^{2}\alpha ^{2}}\int dx\cos [\sqrt{4\pi }
(\varphi _{1}-\varphi _{2})+2(k_{1F}-k_{2F})x]  \label{dBS};
\end{equation}
\begin{equation}
U^B_x\to{\tilde U}_x^{B}=-\frac{1}{2\pi }\int dx\Big[(b_{1}+b_{2})\partial
_{x}\varphi _{1}\partial _{x}\varphi _{2}+(b_{1}-b_{2})\partial _{x}\theta
_{1}\partial _{x}\theta _{2}\Big];  \label{d-X}
\end{equation}
\beq
U^C\to{\tilde U}^C=\frac{f_{C}}{2\pi
^{2}\alpha ^{2}}\int dx\cos \sqrt{4\pi }(\theta _{1}-\theta _{2});
\label{nospinCS}\eeq
so that
\beq
{\tilde H}={\tilde H}_0+{\tilde U}^F+{\tilde U}_d^B+{\tilde U}_x^B
+{\tilde U}^C.
\eeq
Amplitude $f_C$ plays now the role of $t_{sp}$ for spinless electrons.
Analysis of potentially \lq\lq dangerous\rq\rq (in a sense of inducing
instabilities) intersubband processes reduces to estimating
the scaling
dimensions of corresponding cosine operators in terms of the
parameters of the harmonic part. In their turn, these parameters are
related to the Fourier components of the electron-electron interaction
potential. As it  turns out, the latter relation is not that
straightforward, and we will clarify this point in the next Section. 

\subsubsection{Galilean invariance and Pauli principle: single-subband
Luttinger liquid}

To begin with, we consider the simplest case when there is no intersubband
interaction and the Hamiltonian is given by the sum of two single-subband
Hamiltonians (\ref{h0spinless}). (As our discussion
is referred now to a single subband, we suppress
temporarily the subband index.) For a given  effective
1D interaction potential $U\left( x\right)$, the
Luttinger-liquid parameters ($K_{{}}$ and $v$) depend on the $q=0$ and $
q=2k_{F}$ Fourier components of $U_{{}}$, as
well on bare Fermi velocity $v_{F}$: 
\begin{eqnarray}
K_{{}} &=&{\cal K}[U_{{}}(0)/v_{F},U(2k_{F})/v_{F}],  \label{luttK} \\
v &=&v_{F}{\cal V}[U_{{}}(0)/v_{F},U(2k_{F})/v_{F}],\label{luttv}
\end{eqnarray}
where ${\cal K}(x,y)$ and ${\cal V}(x,y)$ are some dimensionless functions
of their arguments. Relations (\ref{luttK}),(\ref{luttv}) have to satisfy  
(i) the Pauli
principle and (ii) Galilean invariance. The Pauli principle for spinless
fermions means that for the case of contact interaction, i.e., when 
$U_{{}}(0)=U(2k_{F})$, the system should behave as if there is no interaction
at all. Accordingly, $K=1$ and $v=v_{F\text{ }}$ for this case, or 
\begin{equation}
{\cal K}(x,x)={\cal V}(x,x)=1.  \label{Pauli}
\end{equation}
Galilean invariance stipulates  that $Kv=v_{F}$, or 
\begin{equation}
{\cal K}(x,y){\cal V}(x,y)=1,\forall x,y.  \label{Galilean}
\end{equation}
Physically, condition (\ref{Galilean}) comes about either by requiring that
the shift of the ground state energy due to the motion of a system as a
whole does not depend on the interaction \cite{haldane}, or by requesting
that the dc conductivity of a uniform system\footnotemark[0] \footnotetext[0]{
Here we consider
a uniform Luttinger liquid. The role of reservoirs, to which the wire
is attached to, will be discussed in Sec.\ref{sec:cond}.}
does not depend on interactions
(Peierls theorem) \cite{schulz_lezush,giam2}. (Also, one can use
the interaction-invariance of the persistent current in a ring threaded by
the Aharonov-Bohm flux).

Conventional bosonization of the $g$-ology Hamiltonian 
(see, e.g., review \cite{schulz_lezush}) leads to
\begin{eqnarray}
K &=&\sqrt{\frac{2\pi v_F + g_4 - g_2}{2\pi v_F + g_4 + g_2}},\nonumber\\
v &=&v_F \sqrt{(1 + \frac{g_4 - g_2}{2\pi v_F}) (1 + \frac{g_4 + g_2}{2\pi v_F})}.
\label{g-ology}
\end{eqnarray}
In terms of the Fourier components of the interaction potential, the 
$g$-parameters 
are expressed
as $g_4=U(0)$ (right-right and left-left amplitude) and $g_2=U(0) - U(2k_F)$ 
(right-left amplitude), and Eq.(\ref{g-ology}) gives
\begin{eqnarray}
K &=&\left[ \frac{1+\frac{U(2k_{F})}{2\pi v_{F}}}
{1+\frac{2U(0)-U(2k_{F})}{2\pi v_{F}}}\right]^{1/2},  \nonumber \\
v &=&v_{F}\left[ 1+\frac{U(2k_{F})}{2\pi v_{F}}\right] ^{1/2}\cdot \left[ 1+
\frac{2U(0)-U(2k_{F})}{2\pi v_{F}}\right] ^{1/2}. \eqnum{wrong} \label{wrongKv}
\end{eqnarray}
One can see
that expressions above do not satisfy conditions (\ref{Pauli}),
(\ref{Galilean}).
Indeed, it follows from Eq.~(\ref{wrongKv}) that $v\neq v_F$
for $U(0)=U(2k_{F})$  and that $Kv\neq v_F$
as long as $U(2k_F)\neq 0$. Usually, spinless Luttinger liquid model
does not include  backscattering explicitly.
The rationale for such a simplification is that for spinless particles
in 1D this process is indistinguishable from forward scattering, 
see , e.g., 
Ref.\cite{metzner}. We do not
find this approach satisfactory, as it is clear that the  behavior of the
system should be determined both by forward and backward
amplitudes. Also, correct
expressions for $K$ and $v$ should include both $U(0)$ and $U(2k_{F})$,
otherwise the Pauli principle cannot be satisfied. This argument can also be
re-phrased in terms of direct and exchange contributions to the self-energy
\cite{mahan}.

What did we do wrong to arrive at the Luttinger-liquid model which does not
satisfy two basic physical principles? As one can show by using the  Ward
indentities (conservation laws) for the system of interacting electrons with 
{\em linear} spectrum\cite{metzner}, the problem occurs already at the level
of fermions and is thus not inflicted by some subtleties of bosonization.
Rather, it is a manifestation of an {\em anomaly}, i.e., a violation of the
conservation law caused by regularization, which one is forced to used in a
model with linear and unbound spectrum \cite{metzner}.

One way to deal with this problem is to replace Eqs.(\ref{wrongKv}) by
expressions which do not follow directly from  the original fermion
Hamiltonian with linear spectrum, but do satisfy all necessary criteria.
This is an accord with the point of view \cite{emery_review} that one should
consider $K$ and $v$ as phenomenological parameters, which are renormalized
from their bare values by irrelevant or marginal operators neglected in
the course of linearization. It turns out that one can find {\em exact}
expressions for $K$ and $v$, satisfying a minimal set of requirements.

First, we notice that the Pauli principle requires ${\cal K}(x,y)$ to be a
function of either $x-y$ or $x/y$. The latter choice contradicts to
the requirement that ${\cal K}$
must have Taylor expansions both around $x=0$ and $y=0$.
Therefore,
\begin{eqnarray}
{\cal K}(x,y) &=&\kappa (x-y), \\
\kappa (0) &=&1.
\label{Pauli1}\end{eqnarray}
Then we notice that the model with forward
scattering only, i.e., the original Luttinger model \cite{luttinger},
 does respect Galilean
invariance. Therefore one can take the Luttinger model 
expression for $K$  as
the correct one, which means that  
\begin{equation}
{\cal K}(x,0)=1/\sqrt{1+x}.
\label{GalileanLutt}\end{equation}
Combining Eq.~(\ref{Pauli1}) with  Eq.~(\ref{GalileanLutt}), we see that
\begin{eqnarray}
{\cal K}(x,y) &=&1/\sqrt{1+x-y}, \\
{\cal V}(x,y) &=&\sqrt{1+x-y},
\end{eqnarray}
or
\begin{eqnarray}
K &=&\left[ 1+\frac{U(0)-U(2k_{F})}{\pi v_{F}}\right] ^{-1/2},
\label{Kright} \\
v &=&v_{F}\left[ 1+\frac{U(0)-U(2k_{F})}{\pi v_{F}}\right] ^{1/2}
\label{vright}.
\end{eqnarray}
The physical meaning of Eqs.~(\ref{Kright}),(\ref{vright}) is
obvious: the effective interaction is equal to backscattering minus
 forward scattering.
One can check that a Luttinger-liquid model with parameters given by
 Eqs.(\ref{Kright}),(\ref{vright}) 
reproduces correctly results for a 1D electron system,
obtained without linearization but in the
limit of weak interactions. For instance, the (inverse) compressibility of a
Luttinger liquid, parametrized by $K$ and $v$ from (\ref{Kright},\ref{vright}),
 is given by
\begin{equation}
\frac{1}{\chi }=\frac{\pi v}{K}=\pi v_{F}+U(0)-U(2k_{F}).  \label{invcompr}
\end{equation}
As one can check, Eq.(\ref{invcompr}) coincides with the inverse
compressibility of electrons with a {\em quadratic } spectrum obtained in the
Hartree-Fock approximation. A perturbative (linear in U) form 
of Eqs.(\ref{Kright},\ref{vright}) 
has recently been derived in \cite{capponi}.
It can also be read off from the tunneling
exponent of a 1D system with a {\it quadratic} spectrum
\cite{matveev93}.

\subsubsection{ Galilean invariance and Pauli principle: coupled subbands}

Now we allow for harmonic coupling between subbands, i.e., take into
account intersubband forward [Eq.(\ref{f-spinless})] and
exchange backscattering [Eq.(\ref{d-X})]. For the contact interaction, the
amplitudes of these two processes coincide:  $b_{1}=b_{2}=f_{0}$ and, as a
result, inter-subband interaction drops out. The Pauli principle
is thus satisfied. Intersubband exchange backscattering does violate
Galilean invariance, and the correction procedure, similar to that for a
single subband, is necessary. We will not do it here  however, because for a
long-range interaction ($a/d\ll 1$), the violation is \lq \lq weak\rq\rq : 
the
deviation from the Galilean-invariant result is proportional to the {\em 
exchange} amplitudes, which are small compared to the {\em direct }ones.

\subsection{Nearly equivalent subbands}
\label{near-spinless}

First, we discuss the CDW-instability, which may occur if 
the density equilibration between subbands is energetically
favorable. To simplify the discussion, we consider the case
of a long range interaction ($a/d\ll 1$), when amplitudes
of exchange processes are small. In the leading order,
$f_C=b_{1,2}=0$ and
the only \lq\lq dangerous\rq\rq\/ process to be considered
is direct intersubband backscattering. Furthermore,
we assume that subbands are nearly equivalent and put $v_{1F}=v_{2F}$ 
and $K_1=K_2$ but keep $\delta k_F=
k_{1F}-k_{2F}$ in Eq.~(\ref{dBS}) finite.
It is convenient to introduce symmetric and antisymmetric
combinations of boson fields
\beq
\varphi_{\pm}=\frac{\varphi_1\pm\varphi_2}{\sqrt{2}};\,
\theta_{\pm}=\frac{\theta_1\pm\theta_2}{\sqrt{2}},
\label{pm}\eeq
which correspond to fluctuations of total (+) and relative
(-) subband charge and current. In terms of these fields,
\begin{eqnarray}
H&=&H_{+}+H_{-},\nonumber\\
H_{+}&=&\frac{1}{2}\int dx \left\{ \frac{
v_{+}}{K_{+}}(\partial_x \varphi_{+})^2+v_{+}K_{+}(\partial_x \theta_{+})^2 
\right\},  \label{H+} \\
H_{-}&=&\frac{1}{2}\int dx
 \left\{ \frac{
v_{-}}{K_{-}}(\partial_x \varphi_{-})^2+
\ v_{-}K_{-}(\partial_x \theta_{-})^2 +\frac{f_{bs}}{\pi^2 \alpha^2} \cos[
\sqrt{8\pi}\varphi_{-} + 2\delta k_F x] \right\},  \label{H-}
\end{eqnarray}
where
\beq
K_{+}=\left[1 + \frac{2U(0)-U(2k_F)}{\pi v_F}\right]^{-1/2};\quad
K_{-}=\left[1 - \frac{U(2k_F)}{\pi v_F}\right]^{-1/2},\label{k-},
\eeq
and $v_{\pm}=v_F/K_{\pm}$.
Note that $K_{-} > 1$ for $U(2k_F)>0$, which 
signals effective attraction in the $(-)$ channel.\\

\subsubsection{Collective adjustment of densities as
a commensurate-incommensurate transition}
\label{cdw-case}
To understand how the CDW-instability works, we consider
 first a model situation of $K_{-} < 1$, so that operator
$\cos[\sqrt{8\pi}\phi_{-}]$   is relevant 
 in the RG sense.  Finite   $\delta k_F$  stops the RG-flow at scale
$\ell\sim 1/\ln|\delta k_F|\alpha$, thus precluding
 the system from reaching its strong-coupling limit.  However, this
consideration
 does not take into account the possibility of a collective density
readjustment between
 subbands.  Such a readjustment may occur, if the kinetic energy loss
$\Omega=v_F \delta k_F$  is compensated by
 the gain in the potential energy due to opening of the gap in the (-)
channel.
  In other words, when the difference in electron densities is
sufficiently small,
 the total energy is minimized by equating the densities and opening the
charge gap.
  Obviously, such a process cannot be considered  at the level of
single-particle
 description of transverse quantization.  Instead, one should now treat
 eigenstates and eigenenergies of the wire as being determined by a
self-consistent procedure,
 involving both single-particle and many-body effects.

The mechanism described above can be considered as a
commensurate-incommensurate transition.
  The incommensurability, defined as   ${\cal I}=L^{-1}\langle \int dx
\partial_x
\varphi_{-} \rangle$, where $L$ is the length of the wire, is known
 to have a threshold behavior \cite{pokr,c-ic}:  ${\cal I} \sim
\sqrt{\Omega^2 - \Omega_c^2}%
\Theta(\Omega- \Omega_c)$, where $\Omega_c =
\sqrt{2\pi K_{-}} \Delta_{CDW}$ and the expression for the gap follows from mapping
on the exactly solvable model of a Heisenberg spin chain \cite{luther},
$\Delta_{CDW} \sim (f_{bs})^{1/2(1 -K_{-})}$.
 As follows from the definition of the incommensurability,
 ${\cal I}=0$ implies  $\delta k_F =0$, i.e., equal subband densities. 
 Therefore, the re-adjustment takes place if $\Omega < \Omega_c$.
  Backscattering is then enabled and relevant  (for $K_{-} < 1$), even for
a non-zero initial value of
$\delta k_F$.

What is the physical meaning of this instability?  A simple picture can be
obtained in the
 limit of strong (both inter- and intra-subband) interactions, when the
potential energy
 dominates over quantum fluctuations.  In this case, electrons of each of
the subbands
 form a regular lattice (Wigner crystal).  Boson fields  $\varphi_n$   also
have periodic
 structures with period equal to  $\sqrt{\pi}$  (recall that a shift of
$\sqrt{\pi}$ corresponds
 to adding one electron to the system).  For $f_{bs}>0$,  the energy of
intrasubband
 repulsion
\beq
 f_{bs}\cos[\sqrt{8\pi}\varphi_{-}] = - f_{bs}\cos[%
\sqrt{4\pi}(\varphi_1 -\varphi_2 + \sqrt{\pi}/2)]
\eeq
is minimized by a relative phase shift of $\sqrt{\pi}/2$  between the
subbands,
 which corresponds to a shift of electron lattices by half-a-period.
This is an inter-subband charge-density wave (CDW).


\begin{figure}[t]
\hspace*{5cm}
\epsfxsize=5cm
\epsfbox{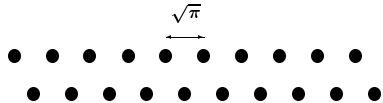}
\vspace{5mm}

\caption{An illustration of the charge density wave in two coupled
subbands. A staggered configuration lowers the energy due to short-range
repulsion, if the densities are commensurate.}
\label{fig:cdw}
\end{figure}

\subsubsection{Competition between CDW and Cooper channels}
\label{dbs-cooper}

Let us now suppose that the density re-adjustment 
did occur, i.e., $\delta k_F=0$, but Cooper scattering is also
present, so that the Hamiltonian of the (-)-channel is
\bea
H_{-}&=&\frac{1}{2}\int dx
 \left\{ \frac{
v_{-}}{K_{-}}(\partial_x \varphi_{-})^2+
\ v_{-}K_{-}(\partial_x \theta_{-})^2 +\frac{f_{bs}}{\pi^2 \alpha^2}
 \cos
\sqrt{8\pi}\varphi_{-} +\frac{f_C}{2\pi^2\alpha^2}\cos\sqrt{8\pi}
\theta_{-}\right\}.  \label{H-c}
\end{eqnarray}
Which of the two instabilities--CDW or superconductivity--wins?
The situation of this type, when cosines of both mutually conjugated
fields ($\varphi_{-}$ and $\theta_{-}$) are present, was analyzed
by Schulz and Giamarchi \cite{giam}. They found that the result
is very sensitive not only to the value of $K_{-}$, which
determines the scaling dimensions of the fields, but also
to the ratio of amplitudes, $f_C/f_{bs}$. As $K_{-} >1$ for repulsive
$U({\bf r})$, it may seem that 
superconductivity
is favored over CDW. The situation is not that straightforward, however.
For example, consider the situation of weak and long-range interactions,
i.e., assume that $U(0),U(2k_F)\ll v_F$ and $a\ll d$. Because
the interaction is weak, both processes are almost marginal, CDW
being on the slightly irrelevant and superconductivity on the slightly
relevant side. For long-range interactions,
$f_{bs}\sim U(2k_F)$ and $f_C\sim (a/d)^2U(0)$.
Modeling $U(x)$ by
\bea
U(x)=\left\{
\begin{array}{ll}
e^2/\epsilon x,\,{\rm for}\, x<d; \nonumber\\
0,\,{\rm for}\,x>d,
\end{array}
\right.
\eea
we get $U(2k_F)/U(0)\sim \ln (k_F a)/\ln (d/a)$. Thus
\beq
\frac{f_C}{f_{bs}}\sim \left(\frac{a}{d}\right)^2\frac{\ln d/a}{\ln k_Fa}\ll 1.
\eeq
The RG-equation for $K_{-}$ \cite{giam}
\beq
\frac{d}{d l}K_{-} = f_C^2 - f_{bs}^2
\eeq
shows that $K_{-}$ decreases, if $|f_C|<|f_{bs}|$.
Even if initially $K_{-}(0)>1$,
the situation with $K_{-}(l)<1$, when CDW is relevant, will be reached 
in the process of renormalization. For weak and long-ranged interactions,
CDW thus wins over superconductivity.

If interactions are not sufficiently weak and/or long-ranged,
only a full RG solution can determine the leading
instability. We will not analyze the general case here. 

\subsection{Non-equivalent subbands: Renormalization Group}

Now we consider a generic situation of non-equivalent subbands, when
$\delta k_F$ is not small enough for the density re-adjustment to occur.
We find that a strong imbalance between Fermi-velocities of occupied subbands
actually helps superconducting instability to develop (see {\it case B} below),
despite the fact that a naive scaling dimension estimate does not show this.
This effect follows from the next-to-leading order perturbative RG 
calculations,
which we present here.

Because $\delta k_F \neq 0$, we
 neglect the dBS process [Eq.~(\ref{dBS})]
from the outset but keep the Cooper one [Eq.~(\ref{nospinCS})].
For long-range interaction, one can also
neglect the xBS process, Eq.(\ref{d-X}), whose amplitude is small
for this case: $b_i
\propto (a/d)^2$. Its inclusion is straightforward ($U_x^{B}$ is
quadratic) but does not lead to any qualitatively new results,
while complicating the analysis
significantly. The 
Hamiltonian then reads
\beq
{\tilde H}={\tilde H}_0+{\tilde U}^F+{\tilde U}^C.
\eeq
Because ${\tilde U}^C$ contains $\theta$-fields, it is convenient
to switch from the Hamiltonian to the Lagrangian approach and to 
integrate out the $\varphi$-fields. The quadratic part of the resulting
 action is diagonalized by the following transformation
\begin{eqnarray}
\left( 
\begin{array}{c}
{\bar\theta}_{1} \\ 
{\bar\theta}_{2}
\end{array}
\right)
 =\left( 
\begin{array}{cc}
\mu_1~~0 &\\
0~~\mu_2 &
\end{array}
\right)
\left( 
\begin{array}{cc}
\cos {\beta }~~\sin {\beta } &  \\ 
-\sin {\beta }~~\cos {\beta } & 
\end{array}
\right) \left( 
\begin{array}{c}
\sqrt{v_1K_1}\theta _{1} \\ 
\sqrt{v_2K_2}\theta_{2}
\end{array}
\right),
\label{unitary}\end{eqnarray}
where
\begin{equation}
\tan {2\beta }=\frac{u_{0}^2}{v_{1}^{2}-v_{2}^{2}},~~
u_{0}=\left[2f_{0}\sqrt{v_{1}K_{1}v_{2}K_{2}}/\pi\right]^{1/2},
\eeq
and
\begin{eqnarray}
&&\mu _{1}=\frac{\cos {\beta }}{\sqrt{v_{1}K_{1}}}-\frac{\sin {\beta }}{%
\sqrt{v_{2}K_{2}}},  \nonumber \\
&&\mu _{2}=\frac{\sin {\beta }}{\sqrt{v_{1}K_{1}}}+\frac{\cos {\beta }}{%
\sqrt{v_{2}K_{2}}}.  \label{mu}
\end{eqnarray}
In terms of new fields, the action is given by
\begin{equation}
S=\frac{1}{2}\int dxd\tau \left[ \sum_{n}R_{n}\{\frac{1}{u_{n}}(\partial
_{\tau }\bar{\theta}_{n})^{2}+u_{n}(\partial _{x}\bar{\theta}_{n})^{2}\}+%
\frac{f_{C}}{\pi ^{2}\alpha ^{2}}\cos \sqrt{4\pi }(\bar{\theta}_{1}-\bar{%
\theta}_{2})\right],  \label{S_spinless}
\end{equation}
where 
\begin{equation}
u_{n}^{2}=\frac{1}{2}\Big(v_{1}^{2}+v_{2}^{2}\pm \sqrt{%
(v_{1}^{2}-v_{2}^{2})^{2}+u_{0}^{4}}\Big)  \label{velocity}
\end{equation}
are the velocities of new collective modes 
and  $R_{n}=1/(u_{n}\mu _{n}^{2})$ are the new stiffness coefficients.

We are now ready to perform the momentum-shell 
RG, i.e., to expand
perturbatively in coupling constant $f_C$ and integrate out high-energy
modes with 2-momentum $k$ within a thin strip $\Lambda - d\Lambda \leq
k \leq \Lambda$ ($d\Lambda/\Lambda \ll 1$). The first-order contributions
renormalize $f_C$, whereas
the second-order ones renormalize stiffnesses $R_n$. The main difference
between our case and the conventional RG-treatment of the sine-Gordon action
(see, e.g., Ref.~\cite{kogut}) is that
the $f_C^2$-contribution produces (among others) mixed gradient terms of
the type $\partial_\nu{\bar\theta}_1\partial_\nu{\bar\theta}_2$ ($\nu=\tau,x$),
which are absent in bare action (\ref{S_spinless}).
To eliminate these terms, we transform
 fields one more time :
\bea
\left(
\begin{array}{c}
{\tilde \theta}_1\\
{\tilde \theta}_2
\end{array}
\right)=\left(1+\frac{\beta'}{t'}\right)
\left(
\begin{array}{cc}
\cosh{\beta^{\prime}} ~~t^{\prime}\sinh{\beta^{\prime}} &  \\ 
\frac{1}{t^{\prime}}\sinh{\beta^{\prime}} ~~\cosh{\beta^{\prime}} & 
\end{array}
\right)
\left(
\begin{array}{c}
{\bar \theta}_1\\
{\bar \theta}_2
\end{array}
\right),
\eea
where 
\beq
t^{\prime}=\frac{u_2 R_2}{u_1 R_1}\,{\rm ~and~}\, \beta^{\prime}=
\frac{d\Lambda}{\Lambda}\left(\frac{f_C}{\pi}\right)^2\frac{1}{u_1u_2R_1R_2}
\eeq
are chosen in such a way that the coefficients in front of mixed
gradient terms vanish.
When  written in terms of ${\tilde\theta}_n$,
the action is brought into its original form but with renormalized parameters.
Resulting RG
equations read 
\begin{eqnarray}
\frac{d}{dl}\frac{1}{R_1}=&&-\bar{f}^2 \frac{1}{R_1}\Big(\frac{4\gamma^2}{
R_2^2 (1 + \gamma^2)} + \frac{2}{R_1^2 (1 + \gamma^2)} + \frac{\gamma}{R_1
R_2}\Big),\label{rg1} \\
\frac{d}{dl}\frac{1}{R_2}=&&-\bar{f}^2 \frac{1}{R_2}\Big(\frac{4}{ R_1^2 (1
+ \gamma^2)} + \frac{2\gamma^2}{R_2^2 (1 + \gamma^2)} + \frac{1}{R_1 R_2
\gamma}\Big),\label{rg2} \\
\frac{d}{dl}\gamma=&&\bar{f}^2\frac{1-\gamma^2}{R_1 R_2},\label{rg3} \\
\frac{d}{dl}\bar{f}=&&\Big(2 - \frac{1}{R_1} -\frac{1}{R_2}\Big)\bar{f},
\label{rg4}
\end{eqnarray}
where $\bar{f}=(1/\pi)f_C\sqrt{u_1^{-2}+u_2^{-2}}$
is the dimensionless coupling constant and $
\gamma=u_1/u_2$.
To the $\bar{f}^2$-accuracy, all terms multiplying $\bar{f}^2$
on the right-hand-side of the first three equations above
have to be treated as 
constants determined by the initial conditions. The system of RG-equations
has
an obvious integral of motion 
\begin{equation}
C=\frac{x^2}{c_1}+\frac{y^2}{c_2}-\bar{f}^2 , \label{integral}
\end{equation}
where
\beq 
x\equiv1-1/R_1,\quad y\equiv 1 -1/R_2,
\eeq
 and $c_{1,2}$ are
the coefficients in front of $\bar{f}^2$ in Eqs.~(\ref{rg1},\ref{rg2}), respectively.
Note also that $x=\left(c_1/c_2\right)y + p$ (we denote $x=x(l),~y=y(l)$,
whereas initial values are denoted by sub-index $0$, i.e., $x(0)=x_0$,
etc.). Constants of motion $C$ and $p$ are determined by initial
conditions.

The flow described by (\ref{rg1}-\ref{rg4}) is quite similar to that of
a canonical Kosterlitz-Thouless system: $x$ and $y$ increase  with $\bar{f}$
regardless of its sign. If $x_0, ~y_0 > 0$,
$\bar{f}$ grows unrestrictedly, flowing into the strong-coupling regime with
a gap in the $\theta_1 - \theta_2$ channel.
Such initial conditions correspond
to $R_{1,2} > 1$, i.e., to the attractive interaction in the $\bar{\theta}
_n$-channels. It is worth emphasizing here that due to the presence of
inter-subband forward scattering, such effective attraction may arise
in a purely repulsive system, as we shall demonstrate shortly ($case$ $A$
below). Another relevant limit is represented by the \lq\lq repulsive\rq\rq\/
 case ($
case$ $B$), where initially $x_0 < 0$, $y_0 < 0$. For a strong repulsion ($%
R_{1,2} \ll 1$), $\bar{f}$ quickly renormalizes to zero 
and the resulting phase is a two-subband Luttinger liquid.
However, there is a region of anomalously small $x_0, y_0 \sim 
\bar{f}_0$ (which requires strong inter-subband scattering), where Cooper
scattering may still be important. One finds that if 
\begin{equation}
(c_1 + c_2) \bar{f}_0^2 > (x_0 + y_0)^2  \label{condition}
\end{equation}
The Cooper process wins over repulsion and initially negative variables
$x, y$ change sign during renormalization. In this case $f_c$  decreases with $l$, initially but then passes through
a minimum and flows finally into strong-coupling regime $\bar{f} \geq 1$.
Equation (\ref{condition}) is a condition for the development of superconducting
fluctuations in the system with purely repulsive interactions.

Now we again apply our analysis to a wire with weak and long-range
interactions. Two limits are possible.\\

{\it {Case A:}} $\Delta_{CDW}/k_F \ll v_{1F}-v_{2F} \ll U_0$.

The first inequality allows one to neglect direct backscattering, 
which leads to inter-subband CDW,
 whereas the second one
allows to consider subbands as \lq\lq nearly equivalent\rq\rq\/.
Denoting $v_F \equiv v_{1F}$, $\delta v_F \equiv v_{1F}-v_{2F}$,
and the  $2k_{F}$-component of the
interaction potential in the $n$-th subband by $U_{2k_F}^{(n)}$,
one finds
\begin{equation}
\frac{1}{R_1} + \frac{1}{R_2}=2 - \frac{U_{2k_F}^{(1)} + U_{2k_F}^{(2)}} {%
2\pi v_F} -\frac{\delta v_F}{v_F}\frac{U_{2k_F}^{(1)}}{2\pi v_F} \leq 2,
\label{casea}
\end{equation}
which corresponds to effective attraction. Thus, thanks to inter-subband
forward scattering, the Cooper process is relevant in the system with
purely repulsive interaction. 
Note that a small velocity imbalance $\delta v_F > 0$
enhances the relevance of Cooper scattering .
 (As our second subband is chosen to have a higher
energy, $\delta v_F$ is always positive.) \\

{\it Case B:} $v_{1F}-v_{2F} \gg U(0)$. {\it Van Hove singularity.}

In this limit an analytic solution is also possible.
Generally, one finds that $1/R_1 + 1/R_2 > 2$, which
corresponds to effective repulsion. Neither CDW nor superconducting
instability can develop, and the resulting phase is a two-subband
Luttinger liquid. This is not true, however, in the limit of a strong velocity
imbalance, when $v_{2F}/v_{1F} \ll 1$, i.e., when the second subband 
just opens for conduction.
Then $u_n \approx v_n$ and $\gamma=u_1/u_2
\sim v_{1F}/v_{2F} \gg 1$. Hence $c_1 \sim \gamma$ and it follows from (\ref{condition}%
) that Cooper process wins over repulsion, if $f_C > U(0)/\sqrt{%
\gamma}$. For long-range interactions, the last inequality reduces to
\begin{equation}
\frac{v_{2F}}{v_{1F}}\ll\left(\frac{a}{d}\right)^4.
\end{equation}
The physics of this scenario is well-known - interactions are
enhanced due to the large value of the density of states ($\propto 1/v_{2F}$)
in the upper subband (Van Hove singularity). We
should warn here that our calculations do not describe the very onset of
conduction in the upper subband, because its proper description requires
accounting for the nonlinearity of the electron spectrum, which is beyond
our bosonization analysis. However, such a calculation was performed by
Balents and Fisher \cite{leon2}, who analyzed the case of a contact
 interaction. They found that superconducting fluctuations
are indeed enhanced in this limit.

We thus see that superconducting
fluctuations do have a good chance to overcome the electron-electron
repulsion and
drive the system into a strong-coupling phase with the gap in the spectrum
of relative current fluctuations, $\bar{\theta}_1 - \bar{\theta}_2$.

There is another important feature of the RG-flow described
by Eqs.~(\ref{rg1}-\ref{rg4}): the interaction tries
to equilibrate densities in the subbands. This is seen from 
equation (\ref{rg3}): $d\gamma/dl$ is proportional to $1-\gamma^2$, which makes
$\gamma=1$ a stable fixed point. $\gamma$ tends to increase, if initially $
\gamma_0 <1$, and it tends to decrease, if $\gamma_0 > 1$.

\section{Electrons with spins}
\label{spinful}

Guided by the results of the previous Section, we now comment briefly
on what happens if spin is included.
As should be clear from the complexity of 
Eqs.(\ref{charge-density}-\ref{singlet-triplet}), 
this question has no simple answer.
For a quantum wire with $0 < U(2k_F) < U(0)$, possible phases are again
(i) Luttinger liquid, (ii) inter-subband CDW, and (iii) Cooper phase 
(superconductor).
As with spinless electrons, 
subbands must be nearly equivalent in order for the CDW phase
to occur,
whereas the Cooper phase needs effective attraction in the relative
charge-density excitation channel. When neither of these conditions
is met, a two-subband Luttinger liquid is realized.
At different degrees of generality, renormalization group analysis
of the model defined by Eqs.(\ref{charge-density}-
\ref{singlet-triplet}) has been performed in the past and we refer
to papers \cite{varma,fabrizio,leon2,ekz} for a detailed description.

\subsubsection{Long-range interactions}

If the interaction is long-range and weak,
a considerable simplification occurs. In this case, amplitudes of
forward intra- and inter-subband processes are the same
[see discussion after Eq.(\ref{cooper})] and a simple perturbative
estimate of scaling dimension $\delta_C$ of the Cooper process 
(\ref{singlet-triplet}) is possible.

For weak interactions, $K_{n\rho}=1-(2U(0) -
U_{2k_F}^{(n)})/2\pi v_{nF}$, and $f_0=U(0)$. 
The $SU(2)$-invariance requires that $K_{n\sigma}=1$.
One finds
\begin{equation}
\delta_{C}=2 - \frac{U_{2k_F}^{(1)} + U_{2k_F}^{(2)}}{4\pi v_F} -\frac{%
\delta v_F}{v_F}\frac{U_{2k_F}^{(1)}}{4\pi v_F} \leq 2.  \label{i}
\end{equation}
Observe that this result coincides with Eq.(\ref{casea}) upon
replacing $U(0) \rightarrow 2U(0)$.
Thus, Cooper scattering is  
relevant for repulsive long-range interactions
(and assuming also that $0 < \delta v_F \ll U_0$).

However, if $K_{\rho-}$ is sufficiently close to its non-interacting
value, i.e., to unity, backscattering is strong ($f_{bs} \gg t_{sp}, t_{tp}$),
and $\delta k_F$ is small,
the CDW-channel can take over the Cooper one, similarly
to the scenario described in Sec.~\ref{dbs-cooper}.

\subsubsection{Electron ladder}
\label{sec:ladder}

There is an important question where an RG consideration
is very helpful.
Suppose that condition  (\ref{i}) is satisfied and thus tunneling
of fluctuational Cooper pairs is relevant.
What happens to spin excitations?
To answer this question, we relax the $SU(2)$-invariance condition
and perform the RG calculation for two
nearly equivalent subbands so that
$v_{n\nu }\equiv v_{\nu },K_{n\nu }\equiv K_{\nu}$, where
$\nu =\rho ,\sigma$.  Nevertheless, we assume that
$\delta k_F$ is still finite and neglect direct backscattering 
(\ref{dir_backward}), similarly to Subsection \ref{near-spinless}.
The problem then becomes identical to that of two coupled
equivalent chains coupled by the interaction (\lq\lq electron ladder\rq\rq\/).
Also, for the sake of simplicity, we consider only the singlet channel
of Cooper scattering.
Due to enhanced symmetry, the total
current fluctuation mode $\theta _{\rho +}=
(\theta _{1\rho }+\theta _{2\rho })/\sqrt{2}$ 
decouples from the rest 
and is described by a harmonic action
 with 
\beq
K_{\rho +}=\left[K_{\rho }^{-2}+
2f_{0}/\pi v_{F}\right]^{-1/2}.
\eeq 
Relative current fluctuations $\theta _{\rho-}=
(\theta _{1\rho }-\theta _{2\rho })/\sqrt{2}$
are described by
the sine-Gordon theory [$t_{sp}$ term in (\ref{singlet-triplet})] 
with 
\beq
K_{\rho -}^{-2}=\left[K_{\rho }^{-2}-
2f_{0}/(\pi v_{F})\right]^{-1/2}\quad{\rm and}\quad 
v_{\rho -}=v_{\rho }[1-2f_{0}K_{\rho }/\pi v_{F}]^{1/2}.
\eeq 
In addition, we have to keep track of the
spin-density sector, which contain the cosine
term corresponding to intra-subband backscattering (\ref{spin-density}).
For convenience, we denote the amplitude
of this term by  $g_{\sigma}$, its initial value
being $g_{\sigma}(0)=U(2k_F)$. After tedious but
straightforward calculations we arrive at the following 
system of RG equations:
\begin{eqnarray}
&&\frac{d}{d\ell}\bar{g}=2(1-K_{\sigma })\bar{g}-\bar{t}^{2},  \label{g-eqn} \\
&&\frac{d}{d\ell}\bar{t}=(2-K_{\sigma }-\frac{1}{K_{\rho -}} - \bar{g})\bar{t},
\label{t-eqn} \\
&&\frac{d}{d\ell}(1-K_{\sigma })=\frac{1}{2}(\bar{g}^{2}+\bar{t}^{2}), \\
&&\frac{d}{d\ell}(1-\frac{1}{K_{\rho -}})=\bar{t}^{2},  \label{rg-system2}
\end{eqnarray}
where $\bar{g}=g_{\sigma }/(\pi v_{F})$ and 
$\bar{t}=t_{sp}/(\pi v_F)$.
Let us recall what happens in the absence of Cooper tunneling first
and set $t_{sp}=0$ everywhere in this system. In the weak-coupling
limit, $K_{\sigma}=(1 - \bar{g})^{-1/2}\approx 1 + {\bar g}/2$ and 
(\ref{g-eqn}) becomes
$d\bar{g}/d\ell=-\bar{g}^2$, which gives 
$\bar{g}_{\ell}=\bar{g}_0/(1 + \bar{g}_0 \ell)$.
For repulsive interactions ($\bar{g}_0 > 0$), $\bar
g\propto\ell^{-1}\to 0$ as $\ell \rightarrow \infty$: 
intra-subband backscattering is marginally irrelevant.
Observe now that when the Cooper process is present and relevant,
i.e., when ${\bar t}$ increases, the flow of ${\bar g}$ is modified:
the ${\bar t}^2$-term on the right-hand-side of Eq.~(\ref{g-eqn}) changes the sign
of $d\bar{g}/d\ell$. Hence
$\bar{g}_{\ell}$ is bound to become negative in the process of renormalization
and grows unrestrictedly in its absolute value. 
Intra-subband spin backscattering is thus driven relevant
by singlet-pair tunneling, which results
in pinning of $\varphi_{\sigma}$ in Eq.(\ref{spin-density})
and opening of the {\it spin gap}. Thus, similar to the true superconducting
state in higher dimensions, the Cooper phase is characterized by gaps
in both charge- and spin-channels. The only massless excitations
are those of the total charge channel.
This phenomenon is not restricted to the degenerate electron
ladder but rather is a generic feature of the system of coupled subbands
and/or chains, see, e.g., Refs.~\cite{leon2,ekz}.

\section{Conductance}
\label{sec:cond}

Having realized the importance of inter-subband interactions, we now proceed
with the analysis of its effect on observable properties of quantum wires. The
first property we consider is conductance $G$. 
\subsection{No disorder}
\label{sec:nodisorder}
Our results for the conductance of a clean wire can be understood from
the following simple considerations. The dc conductance of a single-subband
wire is equal to $2e^2/h$ regardless of the interactions in the wire
\cite{safi,maslov-stone,ponomarenko}. Consider now a wire with
several subbands occupied.
Those interband interactions, which do not open gaps, lead only to a renormalization
of Luttinger-liquid parameters. As these parameters do not enter the final
result for $G$, the conductance
 remains quantized in units of $2e^2/h$ per occupied subband.
Other processes, such as direct backscattering and Cooper scattering, may open
gaps in channels of {\it relative} charge fluctuations as well as in spin
channels. Neither of these gaps, however, affects the {\it center-of-mass}
of the electron fluid through the wire, which continues to move along the wire unrestrictedly. Therefore, $G$ remains unrenormalized 
by this type of interactions as well.

Now we demonstrate the proof of the statements made above. Consider the 
case of a 
superconducting instability, when the cosine term of Cooper scattering
in Eq.~(\ref{cooper}) is relevant and the $\theta_{\rho-}$-field is thus 
gapped.
Gaussian fluctuations of
the gapped field can be described by expanding the relevant
cosine term around its minimum value: 
\beq
\left(4f_C/\pi^2
\alpha^2\right)\cos[\sqrt{2\pi} (\theta_{1\rho}-\theta_{2\rho})]
\approx {\rm const}+m^2
(\theta_{1\rho}-\theta_{2\rho})^2,
\label{semicl}\eeq
where $m$ is the mass of the field.
The strong-coupling (superconducting) phase corresponds to $m\neq 0$, whereas
in the Luttinger-liquid phase $m=0$. 
Expanding $\cos[\sqrt{2\pi}\varphi_{n\sigma}]$ around their minima as well,
we find that at the Gaussian level charge and spin modes decouple again.
Note though that now these are {\it massive} modes.

As spin excitations do not affect charge transport,
we concentrate on the charge sector of the theory, whose 
Hamiltonian is given by the sum of Eqs.~(\ref{charge-density},\ref{forward})
and (\ref{semicl}).
To simplify notations, we suppress index $\rho$ in this section, so
that $\theta_{\rho -} \rightarrow \theta_{-}$, etc.
Using $v_{n} K_{n}=v_{nF}$, we write the
charge Hamiltonian as
\begin{eqnarray}
H_{\rho}=&&\frac{1}{2}\int dx \Big\{ \frac{v_{1F} + v_{2F}}{2} \left[
(\partial_x\theta_{+})^2 + (\partial_x\theta_{-})^2 \right] + \frac{1}{2}\left(
\frac{v_1}{K_1} + \frac{v_2}{K_2}\right) 
\left[(\partial_x\varphi_{+})^2 +
(\partial_x \varphi_{-})^2 \right]  \nonumber \\
&&+\frac{2f_0}{\pi}\left[(\partial_x\varphi_{+})^2 - (\partial_x
\varphi_{-})^2 \right]+ (v_{1F} - v_{2F})\partial_x\theta_{+} \partial_x\theta_{-}
+ \left(\frac{v_1}{K_1} - \frac{v_2}{K_2}\right)\partial_x\varphi_{+} \partial_x
\varphi_{-} + m^2 \theta_{-}^2\Big\}.  \label{Hrho}
\end{eqnarray}
The total charge current is
given by $j=e\sqrt{2/\pi}\sum_n \partial_t \varphi_{n}= e(2/\sqrt{\pi})\partial_t
 \varphi_{+}$. The conductivity
\begin{equation}
\sigma(x,\omega)=\frac{i}{\omega}\int_{-\infty}^{t} dt' \langle 
[j(x,t') j(0,0)] \rangle e^{i\omega t'} = (2e/\sqrt{\pi})^2 (i\omega) 
G_{++}(x,\omega)
\end{equation} 
is related to the retarded Green's function $
G_{++}(x,t)=-i\Theta(t)\langle[\varphi_{+}(x,t), \varphi_{+}(0,0)]\rangle$, which is
coupled to another Green's function
$G_{-+}(x,t)=-i\Theta(t)\langle [\varphi_{-}(x,t), \varphi_{+}(0,0)]\rangle$
by the following equations of motion
\begin{eqnarray}
(-i\partial_t)^2 G_{++}=&&\frac{v_1 + v_2}{2}\delta(x)\delta(t) - \frac{1}{2}%
\partial_x\left\{\frac{v_1^2}{K_1} + \frac{v_2^2}{K_2} + \frac{2f_0}{\pi}(v_1 +
v_2)\right\} \partial_x G_{++}  \nonumber \\
&&-\frac{1}{2}\partial_x\left\{\frac{v_1^2}{K_1} - \frac{v_2^2}{K_2} - \frac{2f_0%
}{\pi}(v_1 - v_2)\right\} \partial_x G_{-+};\nonumber \\
(-i\partial_t)^2 G_{-+}=&&\frac{v_1 - v_2}{2}\delta(x)\delta(t) - \frac{1}{2}%
\partial_x\left\{\frac{v_1^2}{K_1} + \frac{v_2^2}{K_2} - \frac{2f_0}{\pi}(v_1 +
v_2)\right\} \partial_x G_{-+}  \nonumber \\
&&-\frac{1}{2}\partial_x\left\{\frac{v_1^2}{K_1} - \frac{v_2^2}{K_2} + \frac{2f_0%
}{\pi}(v_1 - v_2)\right\} \partial_x G_{++}  \nonumber \\
&&-m^2 \left\{\frac{1}{2}\left(\frac{v_1}{K_1} + \frac{v_2}{K_2}\right) - 
\frac{2f_0}{\pi}\right\}
G_{-+} - m^2 \frac{1}{2}\left(\frac{v_1}{K_1}-\frac{v_2}{K_2}\right) G_{++}~.
\label{conductance-system}
\end{eqnarray}
In the massless limit ($m=0$), this system of equations
is solved readily. In order to model the effect of non-interacting electron 
reservoirs,
which the wire is attached to, we assume that $K_{n},v_n$ vary with $x$ 
adiabatically and approach their non-interacting values of $K_{n}=1, v_n=v_F$ for 
$x\to\pm\infty$
\cite{safi,maslov-stone,ponomarenko}. In the zero-frequency limit,
 the solution is particularly simple: $G_{++}(x,\omega\to 0)=1/
2i\omega,~G_{-+}(x,\omega=0)=0$. As a result, conductivity 
 is $x$-independent, and the
conductance is simply $G=2\times 2e^2/h$. 

In order to see the effect of the gap, we consider first the case of
equivalent subbands (\lq\lq electron ladder\rq\rq\/), 
introduced in Sec.~\ref{sec:ladder}. One observes immediately that
conditions $v_{1}=v_2,~K_{1}=K_2$ lead to complete decoupling of
equations for $G_{++}$ and $G_{-+}$. As a result, the total charge mode $
\varphi_{+}$ is not affected by the gap. Taking the boundary condition for $
K $ and $v$  into account gives again the universal result $G=4e^2/h$.
The result  for the \lq\lq electron ladder\rq\rq\/ thus gives us a hint
 that $G$
 remains
at its universal value despite the presence of the gap in the relative charge
channel. In order to prove this statement in the general case, we
neglect for the moment the boundary conditions for $K_{n}$ and $v_n$, i.e., 
consider
a uniform wire with two coupled subbands. System (\ref{conductance-system}
) is then solved by Fourier transformation. The key feature of the result for $
G_{++}(q,\omega)$ is that it still has a pole corresponding to a massless mode
 $\omega \propto q$, despite the presence of the massive term. The
conductivity becomes
\beq
\sigma(q,\omega) \sim \frac{%
\omega F(\omega,q)}{(\omega^2 - \bar{v}^2 q^2)(\omega^2 + \omega_m^2 -\bar{u}%
^2 q^2)},
\eeq
 where $\bar{v},\bar{u}$ are some renormalized velocities, $
\omega_m$ is some energy proportional to $m^2$, and $F(\omega,q)$ is a
smooth function of its arguments. As a result, 
$\sigma(q,\omega)={\bar G}\delta(q)$
in the limit $\omega\to 0$, where ${\bar G}$ has a meaning of the conductance.
Because we neglected the boundary conditions corresponding to presence
of non-interacting leads in this calculation, 
${\bar G}$ depends on all interaction 
parameters -- $v_{n},K_{n}$, and $m^2$--and is of course different
from $4e^2/h$. It is clear though 
that once the boundary conditions are restored,
this non-universal value is replaced by the universal factor of $4e^2/h$. 
The only other possibility is $
G=0$, which, however, is ruled out by the fact that $G_{++}(q,\omega)$ 
has a massless pole.

We thus conclude that the conductance of a clean wire remains at the universal
quantized value irrespective of
whether the relative charge mode is gapped or not. The case of a CDW
instability can be treated in a similar manner.

\subsection{Disordered wire}

A disordered two-band system in the presence of interaction-induced
instabilities was considered by  Orignac and Giamarchi \cite{orignac}
and by  Egger and Gogolin \cite{gogolin}. Our discussion
of a disordered two-subband wire follows largely these two papers.

Results of the subsequent analysis can be summarized as follows.
If Cooper scattering
opens a gap, the system does not become a real superconductor: 
a single 
weak impurity splits eventually the wire into two disconnected halves 
at low enough energies, and even a 
weak random potential leads to localization of electrons, similar
to the case of a gapless Luttinger liquid. 
Nevertheless, effects of disorder
are less pronounced than for a gapless Luttinger liquid.
On the contrary, the CDW-state
is more sensitive to disorder than a gapless Luttinger liquid.

\subsubsection{Spinless electrons}
\label{disnospin}

We begin by considering a single impurity described as a potential
perturbation
$w(x,\rperp)$. The impurity causes backscattering of electrons
within the occupied subbands, as well as inter-subband
backscattering. The amplitudes of corresponding processes
are given by
\bea
W_n(2k_{nF})&=&\int dx d\rperp w(x,\rperp) \phi_n^2(\rperp) \cos(2k_{nF}x),
\,n=1,2;\label{wndef} \\
W_{\text{inter}}&=& \int dxd\rperp w(x,\rperp) \phi_1(\rperp)\phi_2(\rperp)
\cos\left[(k_{1F}+k_{2F})x\right].\label{winterdef}
\eea
If $w$ varies slowly across the wire, then $W_{\text{inter}}\ll W_n$ due
to the orthogonality of transverse wavefunctions, and we consider
intra-subband backscattering first.
The bosonized form of intra-subband 
backscattering is
\begin{equation}
W_{\text{intra}}^{n}=\frac{W_n(2k_{nF})}{\pi\alpha}\cos[\sqrt{4\pi}\varphi_n]= 
\frac{W_n(2k_{nF})}{\pi\alpha}\cos[\sqrt{2\pi}(\varphi_{+} \pm \varphi_{-})],
~n=1,2; \label{w_intra}
\end{equation}
so the total backscattering operator is given by
\begin{eqnarray}
W_{\text{intra}}=&& \sum_{n=1,2} W_{\text{intra}}^{n}=
\frac{W_2(2k_{2F}) - W_1(2k_{1F})}{\pi \alpha}\sin[\sqrt{2\pi}\varphi_{+}]
\sin[\sqrt{2\pi} \varphi_{-}]
\nonumber\\
&& + \frac{W_2(2k_{2F}) + W_1(2k_{1F})}{\pi \alpha}
\cos[\sqrt{2\pi}\varphi_{+}]\cos[\sqrt{2\pi} \varphi_{-}].
\label{w_intra_2}
\end{eqnarray}
Note that $W_{\text{intra}}$ is local in space and thus 
cannot change the RG-flows
of bulk parameters of the wire. Depending on these bulk parameters, however,
$W_{\text{intra}}$ will either grow, splitting eventually 
the wire into two disconnected
pieces, or decay, in which case the impurity
effectively disappears.

(i) {\it Cooper phase.}\\
\noindent
In the Cooper phase, $\theta_{-}$ is gapped, hence $\varphi_{-}$
fluctuates strongly. [This follows from the fact that $\theta_{-}$ and
$\varphi_{-}$ are canonically conjugated fields, see Sec.~\ref{sec:dos}.]
On the first sight, it may seem that these strong
fluctuations render $W_{\text{intra}}$ to zero. To see that it is not
so, consider
a second-order impurity contribution, e.g.,
\begin{equation}
\left(\frac{W_1(2k_{1F})}{\pi\alpha}\right)^2 \int d\tau \int d\tau^{\prime}\langle e^{i
\sqrt{2\pi}\varphi_{-}(\tau)} e^{-i\sqrt{2\pi}\varphi_{-}(\tau^{\prime})}
\rangle \cos[\sqrt{2\pi}\varphi_{+}(\tau)]\cos[\sqrt{2\pi}\varphi_{+}(\tau^{
\prime})]  \label{t-t'}
\end{equation}
As it will be explained in more details in Sec.~\ref{sec:dos},
the correlator of $\varphi_{-}$--fields in the
Cooper phase decays exponentially, i.e.,
$\langle e^{i\sqrt{2\pi}
\varphi_{-}(\tau)} e^{-i\sqrt{2\pi}\varphi_{-}(\tau^{\prime})} \rangle \sim
e^{-\Delta_{SC} |\tau - \tau^{\prime}|}$, where $\Delta_{SC}$ is the
Cooper gap in the $\theta_{-}$--channel. As a result, the double
integration  over $\tau,\tau'$ in Eq.~(\ref{t-t'}) is effectively
contracted into a single one, the result of integration being
\beq
\Delta_{SC}^{-1}
\left(W(2k_{1F})/\pi\alpha\right)^2 \int d\tau \cos[\sqrt{
8\pi}\varphi_{+}(\tau)].
\eeq
The mechanism of generating higher
order impurity backscattering
was discovered in \cite{orignac,gogolin}.
($\sqrt{8\pi}$ under the cosine indicates that this is a two-particle
backscattering process.)
Following the RG-calculations of Kane and Fisher \cite
{kane_fisher}, one finds that impurity backscattering, generated in this way,
becomes relevant for $K_{+} < 1/2$. Note that this requires rather strong
electron repulsion. For weaker repulsion, backscattering amplitude scales to zero and the
wire retains a universal conductance of $2\times e^2/h$. Without
superconducting correlations, i.e., when $\theta_{-}$ is not gapped,
an impurity is effective for $K < 1$. Thus the Cooper phase weakens but 
does not eliminate
impurity scattering.

(ii) {\it CDW phase.}\\
\noindent
In this phase, $\delta k_F=0$ and $\varphi_{-}$ is pinned by the bulk
nonlinear term $\cos[\sqrt{8\pi}\varphi_{-}]$,
so that $\varphi_{-}$ acquires average value 
$\langle \varphi_{-}\rangle=\sqrt{\pi/8}$.
Allowing for fluctuations around the average value, we
substitute
$\varphi_{-}=\langle \varphi_{-}\rangle + \delta\varphi_{-}$ into
(\ref{w_intra_2}), which gives
\begin{equation}
W_{\text{intra}}=
\frac{W_2(2k_{F}) - W_1(2k_{F})}{\pi \alpha}\sin[\sqrt{2\pi}\varphi_{+}]
\cos[\sqrt{2\pi} \delta\varphi_{-}] - \frac{W_2(2k_{F}) + W_1(2k_{F})}{\pi 
\alpha}
\cos[\sqrt{2\pi}\varphi_{+}]\sin[\sqrt{2\pi}\delta \varphi_{-}]
\label{w_intra_3}
\end{equation}
Observing that for small fluctuations one can replace
$\cos[\sqrt{2\pi} \delta\varphi_{-}] \approx 1$,
we see that the first term in (\ref{w_intra_3}) 
gives the strongest contribution to backscattering,
which is relevant already for $K_{+} < 2$. The second term requires more work.
To the second order in the amplitude of this term,
an expression similar to (\ref{t-t'}) is generated, but now
it involves the following average
\beq
\langle \sin[\sqrt{2\pi}\delta\varphi_{-}(\tau)]
\sin[\sqrt{2\pi}\delta\varphi_{-}(\tau')] \rangle \sim \frac{\Delta_{CDW}}
{\Delta_0}
\sinh[K_0(\Delta_{CDW}|\tau - \tau^{\prime}|)],
\eeq
where $\Delta_{CDW}$ is
the CDW gap, $\Delta_0$ is the ultraviolet energy cutoff,
 and $K_0(x)$ is the modified Bessel function,
[$K_0(x) \sim e^{-x}/\sqrt{x}$
for $x \gg 1$].
This result is due to the fact that in the {\it massive} phase
\beq
\langle e^{i\sqrt{2\pi}\delta\varphi_{-}(\tau)}
e^{\pm i\sqrt{2\pi}\delta\varphi_{-}(\tau')}\rangle \sim
\exp\left[-K_0\left(\Delta_{CDW}/\Delta_0\right) \mp
  K_0(\Delta_{CDW}|\tau - 
\tau^{\prime}|)\right].
\eeq
As a result, correlation functions of $sines$ and
$cosines$ of massive fields are different:
the first ones decay exponentially
with distance, whereas the second ones reach constant values.
Thus the double integral over $\tau, \tau'$ in (\ref{t-t'}) can be reduced
to the single one again, and, similarly to the Cooper-phase case, two-particle
impurity backscattering, relevant for $K_{+} < 1/2$, is generated.

It is also straightforward to analyze the effect of
inter-subband impurity scattering
\begin{equation}
W_{\text{inter}}=\frac{2W_{\text{inter}}(k_{1F}+k_{2F})}{\pi\alpha}
\cos[\sqrt{2\pi}
\varphi_{+}] \cos[\sqrt{2\pi}\theta_{-}].  \label{w_inter}
\end{equation}
Similarly to Eq.(\ref{t-t'}), we have to average over the strongly fluctuating
$\theta_{-}$--field, which generates again the two-particle
backscattering term
$\sim \cos[\sqrt{8\pi}\varphi_{+}]$, relevant for $K_{+}<1/2$.

Hence, the perturbative correction to the conductance of a CDW wire
behaves as
\beq
-\delta G_{CDW} \propto w^2 \epsilon^{K_{+}-2} + \left(w^2/\Delta_{CDW}\right)^2 
\epsilon^{4K_{+}-2},
\label{condcdw}\eeq
where $\epsilon={\rm max}\{T,{\rm bias}\}$ and where we have also
indicated the order of the impurity potential.
Please note that the exponent of the weak-link counterpart of (\ref{condcdw}),
derived in Eq.(\ref{i-v}) of Section \ref{end-LL}, is not related to the
leading exponent $K_{+}-2$ in Eq.~(\ref{condcdw}) by the conventional duality relation \cite{kane_fisher}.
We conjecture that this violation of duality signals phase transition
separating regimes of weak and strong tunneling.

For the Cooper-phase case, $\cos[\sqrt{2\pi}\theta_{-}]$
is replaced by $\sin[\sqrt{2\pi}\delta \theta_{-}]$. Hence, 
$W_{\text{inter}}$
also generates the effective two-particle term
in the second order of perturbation theory (cf. the
analysis of the second term in (\ref{w_intra_3})). The correction
to the conductance is given by
\beq
-\delta G_{SC}\propto \left(w^2/\Delta_{SC}\right)^2\epsilon^{4K_+-2}.
\label{condsc}\eeq
To summarize, the Cooper phase is insensitive to a single impurity
as long as
$K_{+} > 1/2$, whereas the CDW one is stable only for $K_{+} > 2$.

We now turn to the case of a weak random potential produced by many
impurities. To establish the boundary between delocalized and localized
regimes for the case of weak disorder, it suffices to replace
$\epsilon\to 1/L$ in Eqs.~(\ref{condcdw},\ref{condsc})
 and to multiply $\delta G$ by the
total number of impurities \cite{giam_book}, proportional to $L$.
 Depending on whether
$\delta G$ increases or decreases with $L$, the wire is in the
localized or delocalized phase. By doing so, one concludes that a wire
is localized for $K_+<3$, if it is in the CDW phase, and for $K_+<3/4$, if
it is in the Cooper phase. For comparison, a (spinless) 
single-subband Luttinger
liquid is localized for $K<3/2$.   

\subsubsection{Electrons with spins}

The bosonized form of impurity backscattering is given by

\begin{eqnarray}
W_{\text{intra}}^{1(2)} &=&\frac{4W_{1(2)}(2k_{F1,2})}{2\pi \alpha }\cos [\sqrt{
\pi 
}(\varphi _{\rho +}\pm \varphi _{\rho -})] \nonumber\\
&&\times \cos [\sqrt{\pi }(\varphi _{\sigma +}\pm \varphi _{\sigma -})]; \\
W_{\text{inter}} &=&\frac{4W_{\text{inter}}(k_{1F}+k_{2F})}{2\pi \alpha }
\cos [\sqrt{\pi }
(\varphi _{\rho +}+\theta _{\rho -})] \nonumber \\
&&\times \cos [\sqrt{\pi }(\theta _{\sigma -}+\varphi _{\sigma +})],
\end{eqnarray}
where $+(-)$ in the argument of cosines
refers to the 1st (2nd) subband and all operators are evaluated at
the position of the impurity.

In the Cooper phase, the $\theta _{\rho -}$-- and 
$\varphi _{\sigma \pm }$--modes are gapped, whereas the conjugated
modes, i.e., $\varphi _{\rho -}$ and $\theta _{\sigma \pm \text{ }}$, exhibit
strong fluctuations. Integrating out $\varphi _{\rho -}$ and $\varphi
_{\sigma \pm \text{ }}$, we find
\beq
W_{\text{intra}}\sim \Delta
_{SC}^{-1}\left[ W(2k_{1,2~F})\right]^{2}\cos [\sqrt{4\pi }\varphi _{\rho +}],
\eeq
 which is relevant for $K_{\rho +}<1$. The same is true for $W_{\text{inter}}$,
 where strong
fluctuations of $\theta _{\sigma -}$ produce a similar operator.
The correction to the conductance behaves as
\beq
-\delta G_{SC}\propto \left(w/\Delta_{SC}\right)^2\epsilon^{2K_{\rho +}-2},
\eeq
i.e., as if we were dealing with a single-channel Luttinger
liquid, characterized by parameter $K_{\rho +}$, subject to
an effectively reduced impurity potential.
Weak random potential leads to localization for $K_{\rho+}<3/2$.
 
In the CDW-state, the situation is
different. In this case, the $\varphi _{\rho -}$- and $\varphi _{\sigma \pm
} $-modes are gapped \cite{schulz2,orignac},
whereas the $\theta _{\rho -}$- and $\theta _{\sigma
\pm }$- modes fluctuate strongly. As a result, intersubband
backscattering is renormalized into $\cos[\sqrt{4\pi}\varphi_{\rho+}]$, as
in the Cooper phase,
but intra-subband one remains unchanged and is determined by the
dynamics of the only gapless $\varphi_{\rho +}$ mode:
\begin{equation}
W_{\text{intra}}\propto \cos [\sqrt{\pi }\varphi _{\rho +}].
\end{equation}
Pinning of $\varphi _{\rho +}$
at the impurity site leads to the suppression of the
conductance. $W_{\text{intra}}$ is relevant, i.e., an impurity eventually
splits the wire into two disconnected halves for $K_{\rho +}<4$.
The correction to the conductance behaves as
\beq
-\delta G_{CDW}\propto w^2\epsilon^{\frac{K_{\rho+}}{2}-2}.
\eeq
 This is to
be contrasted with the case of a gapless Luttinger liquid, when the impurity
is relevant only for repulsive interactions ($K<1$). This reflects the fact
that a real (gapped) charge-density-wave is pinned stronger than the
fluctuating one (Luttinger liquid). Finally, weak random potential
localizes the CDW-wire for $K_{\rho+}<6$.
It is worth pointing out here that such large values of critical
$K_{\rho+}$, separating localized and extended regimes, imply
 strong effective attraction between charge fluctuations in the $\rho +$ 
channel.
It might well be that  a charge segregation \cite{dzyal}, instead of the CDW instability, will take place for such a strong attraction. 

Thus, for electrons with spin,
the Cooper phase is more stable to impurities than the CDW one,
similar to a spinless case, but
both are unstable in the physically relevant region of $K_{\rho+} < 1$.

\section{Single-particle density of states}

\label{sec:dos}

We now turn to the discussion of tunneling into a quantum wire.
To the leading order in barrier transparency ${\cal T}$,
the differential tunneling conductance is
\beq
G(V)=\frac{dI}{dV}=|{\cal T}|^2\rho_c\rho(eV),
\eeq
where $\rho_c$ is the density of states (DOS) in the contact
(which we assume to be energy-independent), $\rho(\epsilon)$ is the
DOS of the wire, and $V$ is the applied voltage.
When the wire is in the gapless Luttinger-liquid phase,
$\rho(\epsilon)\propto |\epsilon|^\beta$. This behavior
has recently been observed in tunneling into carbon
nanotubes \cite{cobden}. Tunneling into the edge of a fractional
quantum Hall system also exhibits a power-law current-voltage
dependence \cite{chang}, which might be an indication of a chiral
Luttinger-liquid state at the edge. Here, however, the
situation is not that straightforward, and other explanations,
different from a chiral Luttinger liquid,
have also been suggested \cite{shytov,alekseev}.
  
Suppose now that a two-subband quantum wire is in one of the possible
gapped phases, i.e., CDW or Cooper phase. The goal of this Section is to analyze
what would a tunneling experiment show in this case. The answer turns out
to depend crucially on the geometry of the experiment. If the tunneling
contact probes the interior of the wire, the gapped behavior is predicted:
$G(V)=0$ for $eV<\Delta$, where $\Delta$ is the appropriate energy
gap. For $eV>\Delta$ the behavior is non-universal: the threshold
behavior of $\rho$ is determined by gapless charge and spin modes.
More surprisingly, tunneling into the end of the CDW-wire exhibit 
a gapless behavior,
similar to the Luttinger-liquid case. The tunneling exponent though
is different from that for the gapless phase.
\subsection{Tunneling preliminaries}
The local single-particle (or tunneling) density of states is given by 
\begin{equation}
\rho(\omega,x)=-\frac{1}{\pi} Im\{ G_{ret}(\omega,x)\},
\end{equation}
where $G_{ret}(\omega,x)$ is a Fourier transform of retarded
Green's function $G_{ret}(t,x)=-i\Theta(t)\Sigma_s\langle \{\Psi_s(t,x),
\Psi_s^{\dagger}(0,x)
\}\rangle $. Representing the electron of the 
n-th subband as a sum of right- and left-movers and accounting for the
orthogonality of transverse wavefunctions, the Green's function
becomes $G_{ret}(t,x)=\sum_n \left[G_{ret}^{R_n}(t,x) + G_{ret}^{L_n}(t,x)\right]$,
where the summation is over all occupied subbands and $%
G_{ret}^{N_n}(t,x)=-i\Theta(t)\Sigma_s\langle \{N_{ns}(t,x),N_{ns}^{\dagger}(0,x)\}
\rangle$, 
$N=L,R$, see (\ref{bosr},\ref{bosl}). The contribution of the 
off-diagonal terms 
$\sim \langle R_n L_n^{\dagger}\rangle$ is
less singular and is thus neglected. 
$\rho(\omega,x)$ is a
sum of contributions from right and left movers of all occupied subbands. 
To find $\rho(\omega,x)$, it is convenient to calculate first the Matsubara
Green's function 
\begin{equation}
G^R(\tau,x)=-\langle T_{\tau} R_s(\tau,x)R_s^{\dagger}(0,0)\rangle,
\label{thermal}\end{equation}
and then to make the analytic continuation to real frequencies.
Left- and right-moving fermions give identical contributions to $\rho$, thus
the result obtained from (\ref{thermal}) is simply multiplied by a
factor of two at the end.

The key feature of gapped phases in multisubband 1D systems is the co-existence
of gapped and gapless modes, which also makes the calculations to be
slightly less trivial. The single-particle Green's function
under these circumstances has recently been considered in
 Ref.~\cite{voit,wiegmann}, and our analysis
follows largerly these two papers.

\subsection{Warm-up: DOS of a half-filled Hubbard chain}
\label{sec:mott}
To warm up, we consider the simplest system in which
gapped and gapless modes co-exist
-- a single-band Hubbard chain at half-filling. 
In the context of nanostructure physics, such a system is produced by imposing an
artificial periodic potential of period $a_0$ over a quantum wire \cite{kouwenhoven}.
At half-filling, the Fermi-momentum $k_F=\pi/2a_0$ is commensurate with
the reciprocal lattice spacing, which gives rise to Umklapp scattering.
An Umklapp process occurs as simultaneous backscattering of two right- or 
left-moving
electrons, the total momentum transferred to the lattice being 
$\pm 4\times\pi/2a_0=\pm 2\pi/a_0$. 
This process is responsible for opening of the (Mott-Hubbard) gap in the
charge-excitation spectrum. On the other hand, spin excitations remain
gapless, and are described by the $SU(2)$-invariant Luttinger-liquid
Hamiltonian ($K_{\sigma}=1$).

The corresponding Hamiltonian of the charge sector is 
\begin{equation}
H=\frac{1}{2}\int dx \{v_{\rho}K_{\rho}(\partial_x \theta_{\rho})^2 + \frac{%
v_{\rho}}{K_{\rho}}(\partial_x\varphi_{\rho})^2 + g\cos[\sqrt{8\pi}
\varphi_{\rho}] \}, \label{umklapp}
\end{equation}
where the last (cosine) term represents Umklapp scattering.
Substituting Eq.~(\ref{bosr}) into Eq.~(\ref{thermal}), one finds that
 $G^R$
factorizes into a product of
spin and charge parts 
\begin{eqnarray}
&&G^R(\tau,x)=-\frac{sgn(\tau)}{2\pi\alpha} F_{\sigma}(\tau,x)
F_{\rho}(\tau,x);  \nonumber \\
&&F_{\nu}= \langle \exp[i\sqrt{\frac{\pi}{2}}(\varphi_{\nu}(1) -
\varphi_{\nu}(0))] 
\exp[-i\sqrt{\frac{\pi}{2}}(\theta_{\nu}(1) -\theta_{\nu}(0))] \rangle_{\nu};
\nu=\rho,\sigma,\end{eqnarray}
where shorthand notations $1\equiv (\tau,x),~0\equiv (0,0)$ have been used.
$F_\sigma$ is gapless, whereas $F_{\rho}$ contains massive fields. 
\subsubsection{Bosonic calculation}
\label{sec:mottdosboson}
To calculate $
F_{\rho} $ in the bosonic language, we adopt the semiclassical
approximation, i.e., expand the cosine in (\ref{umklapp})
around its minimum to the second order in fluctuations. This is equivalent to
replacing $g\cos[\sqrt{8\pi}\varphi_{\rho}] \rightarrow m^2
\varphi_{\rho}^2$, which defines the mass $m^2\equiv 4\pi g$. Now the averaging
is straightforward:
\begin{equation}
F_{\rho}(\tau,x)= \exp\left\{-\frac{\pi}{2}\int \frac{d^2 \vec{k}}{(2\pi)^2} (1 - 
\cos[\vec{k}\cdot\vec{z}]) \Big(G_{\theta}(k) +G_{\varphi}(k) + 2iG_{\theta}(k)
\frac{k_0 k_1}{vK k_1^2 + m^2}\Big)\right\},  \label{F_rho}
\end{equation}
where $\vec{k}\equiv(k_0,k_1)=(\omega_n,q)$, $\vec{z}\equiv(\tau,x)$ and 
\begin{equation}
G_{\varphi}(\omega_n,q)\equiv\langle
T_\tau\varphi_{\rho}\varphi_{\rho}
\rangle_{\omega_n q}
=\frac{vK}{v^2 q^2 + \omega_n^2 + m^2 vK}.
\end{equation}
The  Green's function of $\theta_\rho$--fields can be written
as
\begin{equation}
G_{\theta}(\omega_n,q)\equiv\langle T_\tau\theta_{\rho}\theta_{\rho}\rangle_{\omega_n q}
=\frac{v}{K}\Big(\frac{1}{v^2 q^2 + \omega_n^2} + \frac{
\omega_n^2} {v^2 q^2}\frac{\bar{m}^2}{(v^2 q^2 + \omega_n^2)(v^2 q^2 + \omega_n^2
+ \bar{m}^2)}\Big)=G_{\theta}^{(1)}(\omega_n,q) + G_{\theta}^{(2)}(\omega_n,q),
\label{G-theta}
\end{equation}
where $\bar{m}^2\equiv m^2 vK$.
The first term in Eq.~(\ref{G-theta}) is just a free Green's function, whereas the second 
one
is present only in the strong-coupling phase and contains a
infrared divergence at $q\to 0$. This divergence is often
explained by the \lq\lq uncertainty principle\rq\rq\/: in a gapped phase, 
a position-like
field ($\varphi_{\rho}$)
acquires an average value, hence its canonical conjugate, momentum-like
field ($\theta_{\rho}$) fluctuates strongly, 
hence the average $\langle \theta_{\rho}\theta_{\rho}\rangle$
diverges.
 Let us analyze the Fourier transform of $G_\theta$ in more details, and define 
\begin{equation}
I(x,\tau)=\int \frac{d^2 \vec{k}}{(2\pi)^2} (1 - \cos[\vec{k}\cdot\vec{z}])
G_{\theta}^{(2)}(\omega,q).  \label{int}
\end{equation}
Changing to polar coordinates $\omega_n=k\cos{\phi},q=k\sin{\phi}, 
\tau=z\cos{\vartheta},x=z\sin{\vartheta}$, we get
\begin{equation}
I(z,\vartheta)=\frac{\bar{m}^2 z^2}{(2\pi)^2}\int_0^{2\pi} d\phi \frac{
\cos^2{\phi}} {\sin^2{\phi}}\cos^2(\phi-\vartheta) \int_0^{\infty} \frac{dy}{y} 
\frac{1 - \cos{y}}{y^2 + \bar{m}^2 z^2 \cos^2(\phi - \vartheta)}.
\end{equation}
The integral over $\phi$ diverges at $\phi=0$, which, if taken literally, means that 
$I=\infty$ and thus $F_{\rho}(\tau,x=0)
=0 $ for any finite $\tau$. However, this divergence is absent at $
\vartheta=\pi/2$, which corresponds to $\tau=0,z=x$. Let us therefore continue the
calculation at this special point. Despite the cancellation of the infrared
divergence, the integral is still controlled by the region of small $\phi$:
$\phi\sim 1/({\bar m}x)\ll 1$.
Expanding $\sin{\phi} \sim \phi$ and extending the limits of angular integration to 
$\pm \infty$, we find that $I(z,\vartheta=\pi/2)=\bar{m}x/4 + O(1/(\bar{m}
x))$ for $\bar{m}x \gg 1$. Collecting regular contributions from $
G_{\varphi} $ and $G_{\theta}^{(1)}$, we find that the equal-time exponential
correlator of $
\theta_{\rho}$--fields is given by 
\begin{equation}
\langle e^{ia\theta_\rho(x,0)} ~e^{-ia\theta_\rho(0,0)} \rangle
=\exp\left[-\frac{a^2}{4\pi K}
\ln(x^2/\alpha^2) - \frac{a^2 }{4K}\bar{m}x\right].  \label{gapped-corr}
\end{equation}
An important feature here is that the
expected exponential decay of this correlator is modified by 
the power-law prefactor, given by the usual Luttinger-liquid correlator.
Symbolically, $\langle e^{ia\theta_{\rho}(x,0)} 
~e^{-ia\theta_{\rho}(0,0)}\rangle_{\bar{m}\neq 0}
=\langle e^{ia\theta_{\rho}(x,0)} ~e^{-ia\theta_{\rho}(0,0)}\rangle_{\bar{m}=0} 
\times ~e^{-\bar{m}x}$.
 One should be careful in using Eq.(\ref{gapped-corr}): Luttinger-liquid 
parameter
$K$, which appears here, should in fact be understood as the 
strong-coupling fixed-point
value, $K^{*}$, which is often unknown.

Fortunately, the fixed-point value of $K$ is known  for a 
half-filled Hubbard chain:
$K^{*}=1/2$. For ${\bar m}x\gg 1$ we then obtain
\begin{eqnarray}
&&\langle e^{i\sqrt{\frac{\pi}{2}}\varphi_{\rho}(x)} ~e^{-i\sqrt{\frac{\pi}{2}}
\varphi_{\rho}(0)}\rangle \to const,  \nonumber \\
&&\langle e^{i\sqrt{\frac{\pi}{2}}\theta_{\rho}(x)} ~e^{-i\sqrt{\frac{\pi}{2}}
\theta_{\rho}(0)}\rangle \sim \exp[-\pi \bar{m}x/4]/ \sqrt{x} , \label{hubbard-1}
\end{eqnarray}
and the full Green's function behaves as 
\beq
G^R(0,x)\propto\frac{\exp[-\pi \bar{m}x/4]}{\sqrt{x}}
\times \frac{1}{\sqrt{x}},
\label{full}\eeq
in agreement with Ref.~\cite{gulacsi}. The second $x^{-1/2}$-factor in 
Eq.~(\ref{full}) is due to gapless spin excitations.

So far, all calculations have been straightforward. Now we would like to argue
that the infrared divergence of $I(z,\vartheta\neq 0)$ is an artifact of the
semiclassical approximation, which ignores degeneracy of $\cos[\sqrt{8\pi}
\varphi]$ with respect to a uniform shift $\varphi \rightarrow \varphi + 
\sqrt{\frac{\pi}{2}}N$ with integer $N$.
The proper theory of both  massive and massless phases should be
Lorentz-invariant.  We thus propose that the correct result,
valid for any $z=\sqrt{x^2 + v^2\tau^2}$, is given by Eq.(\ref{gapped-corr})
where $x$ is replaced by Euclidian distance $z$: $x\rightarrow z$.
Therefore,
\begin{equation}
\langle e^{ia\theta(x,\tau)} ~e^{-ia\theta(0,0)}\rangle =\exp\left[-\frac{a^2}
{4\pi K} 
\ln(\frac{x^2+ v^2\tau^2}{\alpha^2}) - \frac{a^2 }{4K}\bar{m}
\sqrt{x^2 + v^2\tau^2}\right].
\label{proposal}
\end{equation}
Similar arguments in favor of such replacement were given by
Voit \cite{voit}.

We now use Eq.~(\ref{proposal}) to evaluate Eq.~(\ref{F_rho}), 
and find [compare
with (\ref{hubbard-1})] 
\begin{equation}
F_{\rho}(\tau,0) \sim \sqrt{\frac{\alpha}{v|\tau|}} \exp[-\pi \bar{m}v|\tau|/4].
\label{F_rho_2}
\end{equation}
The spin sector average is non-zero and universal (thanks to $K_{\sigma}=1$), 
therefore $F_{\sigma}(\tau,0) \sim \sqrt{\alpha/v|\tau|}$. The corresponding DOS will
be calculated later, see Eq.(\ref{mott-dos}). The correctness of the procedure
described above is verified in the next Section.

\subsubsection{Re-fermionization.}

To check that our proposition makes sense, we now switch gears and derive Eq.~(%
\ref{F_rho_2}) in a completely different way. To this end, we use the
 Luther-Emery refermionization procedure \cite{luther-emery}, which works
for $K=1/2$, i.e, at the fixed point of a half-filled Hubbard chain. 
This procedure
begins with  an innocuous looking transformation 
$\varphi_{\rho}=\varphi/\sqrt{2}
,~\theta_{\rho}= \sqrt{2} \theta$, which changes Umklapp scattering in Eq.(
\ref{umklapp}) into backscattering of some auxiliary particles (solitons):
$\cos(\sqrt{8\pi}\varphi_{\rho})=\cos(\sqrt{4\pi}\varphi)$. Right- and
left-going solitons are defined by 
\begin{equation}
\psi_{\pm}=\frac{1}{\sqrt{2\pi \alpha}}\exp\left\{\pm i\sqrt{\pi} (\varphi \mp
\theta)\right\}.
\end{equation}
In terms of new bosons, the original
fermion operator (\ref{bosr}) becomes 
\begin{equation}
R_s=\frac{e^{-i\pi/8}}{\sqrt{2\pi \alpha}} \exp[is\sqrt{\frac{\pi}{2}}%
(\varphi_{\sigma} - \theta_{\sigma})] ~ \exp(i\frac{\sqrt{\pi}}{2}\varphi)
\exp(-i\sqrt{\pi}\theta).
\end{equation}
It can also be written in terms of solitons 
\begin{equation}
R_s=e^{i\pi/8}\exp[is\sqrt{\frac{\pi}{2}}(\varphi_{\sigma} - \theta_{\sigma})%
] \exp(-i\frac{\sqrt{\pi}}{2}\varphi) ~\psi_{+},
\label{soliton}\end{equation}
where $\varphi$ is expressed in terms of soliton density fluctuations as $
(1/\sqrt{\pi})\partial_x \varphi= :\psi^{\dagger}_{+} \psi_{+} +
\psi^{\dagger}_{-} \psi_{-}:$. The usefulness of these formal manipulations
is based on the fact that 
Hamiltonian (\ref{umklapp}) is quadratic in $massive$ solitons $\psi_{\pm}$,
and the mass
(or the gap $\Delta$) in their spectrum is determined by coupling
constant $g$: $\Delta=g/(2\pi \alpha)$. Due to the presense of the gap, charge
density fluctuations are suppressed, which results in the suppression of 
fluctuations of $\varphi$. Therefore, at energies below the gap
the phase factor $\exp(-i\sqrt{\pi}\varphi/2)$ in Eq.~(\ref{soliton})
 can be replaced by its average value. The charge part
 of the Green's function ($F_{\rho}$) coincides with the Green's
function of massive fermions 
\begin{equation}
F_{\rho}(\tau,x) \sim \int \frac{d\omega_n dq}{(2\pi)^2} e^{i\omega_n\tau +
iqx} \Big(-\frac{i\omega_n + vq}{\omega_n^2 + \Delta^2 + v^2 q^2}\Big).
\label{Frho}
\end{equation}
At $x=0$,
\begin{equation}
F_{\rho}(\tau,0) \sim -\frac{1}{v}\int d\omega_n 
\frac{i\omega_n e^{i\omega \tau}} 
{\sqrt{ \omega_n^2 + \Delta^2}}=
-\frac{1}{v}\partial_{\tau}\int d\omega_n \frac{\cos(\omega_n \tau)}{
\sqrt{\omega_n^2 + \Delta^2}}=\frac{\Delta}{v} sgn(\tau) K_1(\Delta ~|\tau|).
\label{Frho-local}
\end{equation}
Asymptotically, $F_{\rho}(\tau,0) \sim 
(\Delta/v\sqrt{\Delta |\tau|}) e^{-\Delta |\tau|}$, 
in agreement with our earlier proposition (\ref{F_rho_2}).

It follows {from} Eqs.(\ref{Frho})-(\ref{Frho-local}) that upon continuing to
real frequencies $F_{\rho }^{ret}(\omega =i\omega _{n})=-(\omega/
2v)\left(\Delta ^{2}-\omega ^{2}\right)^{-1/2}$. 
Hence, the DOS of massive fermions
is given by 
\begin{equation}
\rho _{\rho }(\omega ,x=0)=\frac{1}{2\pi v}\Theta (\omega -\Delta )\frac{%
\omega }{\sqrt{\omega ^{2}-\Delta ^{2}}}.  \label{massivedos}
\end{equation}
Up to a factor of $1/2$, which is due to the fact that this is the contribution
of right-movers only, the obtained result
is just the DOS of free massive particles with dispersion $
\epsilon (q)=\sqrt{v^2 q^{2}+\Delta ^{2}}$. Since there are no particles above
energy $\Delta $ at zero temperature, there are no interaction corrections
to the density of states as well \cite{sachdev}.  Another way of
deriving this result consists in using the Ising-model representation of $
\varphi ,~\theta $ fields (see Ref.\cite{shelton} for details). 
In this representation, 
\begin{eqnarray}
&&\exp (-i\sqrt{\pi }\theta )=\sigma _{1}\mu _{2}-i\mu _{1}\sigma _{2}, 
\nonumber \\
&&\exp (i\sqrt{\pi }\varphi )=\mu _{1}\mu _{2}+i\sigma _{1}\sigma _{2},
\end{eqnarray}
where $\mu _{i}$ ($\sigma _{i}$) are order (disorder) fields of the $d=2$
Ising model, whose correlators are known.
 At long times, i.e., when $\Delta \tau \gg 1$, 
\begin{eqnarray}
&&\langle \mu _{i}(\tau)\mu _{j}(0)\rangle \sim \delta _{ij}\langle \mu 
\rangle^{2}, 
 \nonumber \\
&&\langle \sigma _{i}(\tau)\sigma _{j}(0)\rangle \sim \delta _{ij}K_{0}
(\Delta |\tau|).
\label{ising}\end{eqnarray}
As a result, $F_{\rho }(\tau )\sim sgn(\tau) K_{0}(\Delta |\tau| )$. 
Because of the
condition $\Delta \tau \gg 1$, there is no discrepancy between 
Eqs.~(\ref{ising}) and 
(\ref{Frho-local}), since the leading asymptotic term of 
$K_{\nu }(x)$ is $\nu $%
-independent. Hence one again finds a square-root singularity in $F_{\rho
}^{ret}(\omega )$ for $\omega -\Delta \ll \Delta $. The correspondence with the
Ising model allows one to estimate neglected terms as $e^{-3\Delta \tau }$.

A more illuminating way to understand the square-root singularity is provided by
the real-time calculation. Starting from Eq.(\ref{Frho-local}), it can be
shown that \cite{tsvelik2} 
\begin{eqnarray}
Im[F_{\rho}^{ret}(\omega)]=&&\frac{\Delta}{4\pi v} \int_{-\infty}^{\infty} dt
e^{i\omega t} \Big(K_1(-i\Delta t) - K_1(i\Delta t)\Big)  \nonumber \\
&&=\frac{\Delta}{2v}\int_{0}^{\infty}dt \sin(\omega t) Y_1(\Delta t)
=\Theta(\omega - \Delta)\frac{\omega}{2v\sqrt{\omega^2 - \Delta^2}},
\label{estim}
\end{eqnarray}
where $Y_1(x)$ is the Bessel function of the second kind. 
Since asymptotically $Y_1(x)\sim \sin(x)/\sqrt{x}$, the origin of the
singularity at $\omega=\Delta$ can be easily understood. For $\omega - \Delta
\ll \Delta$, the integrand of (\ref{estim}) oscillates very slowly, with
period $t_0=2\pi/(\omega - \Delta)$. The integral is thus determined
by long times, $t \approx t_0$, and can be estimated as $\int_0^{t_0}1/
\sqrt{t} \sim \sqrt{t_0}$. We see that the 
threshold behavior of the DOS is determined by 
times much longer than $1/\Delta$, which justifies
our use of the long-time asymptotics of Bessel functions to evaluate the DOS at 
$\omega \approx\Delta$.

\subsubsection{DOS of a physical electron}

To find the density of states of a physical electron, we have to convolute Eq.(%
\ref{massivedos}) with the contribution of the gapless spin mode:
\begin{eqnarray}
\rho(\omega,x=0)=&&\int \frac{d\epsilon_n}{2\pi} F_{\rho}(\epsilon_n,x=0)
F_{\sigma}(\omega_n - \epsilon_n,x=0)|_{\omega_n =-i\omega}  \nonumber \\
&&= \frac{2}{\pi}\int_0^{\omega} d\epsilon Im[F_{\rho}^{ret}(\epsilon)]
Im[F_{\sigma}^{ret}(\omega - \epsilon)]. \label{convolution}
\end{eqnarray}
Since $F_{\sigma}(\epsilon_n) \sim \sqrt{\alpha/iv \epsilon_n}$, we
find 
\begin{equation}
\rho(\omega,x=0)=\frac{2}{\pi v}\sqrt{\frac{\alpha}{v}}\Theta(\omega - \Delta)
\int_{\Delta}^{\omega} d\epsilon \frac{\epsilon} {\sqrt{\epsilon^2 - \Delta^2
}}\frac{1}{\sqrt{\omega - \epsilon}}.  \label{mott-dos}
\end{equation}
For $\omega - \Delta \ll \Delta$, where our derivation is valid,
\beq
\rho(\omega,x=0)=\frac{\pi}{v} \sqrt{\frac{g}{v}} 
~\Theta(\omega - \Delta).
\label{mottdos}\eeq 
Instead of a square-root singularity of 
Eq.~(\ref{massivedos}), the DOS of a physical electron exhibits a regular 
behavior approaching a finite value at the
threshold. This modification is due to dressing of the gapped charge mode
by gapless spin excitations. At energies much above the gap DOS increases,
$\rho(\omega,x=0) \sim \sqrt{\omega}$, which means that the spectral weight
is shifted to higher energies. The energy-independent electron DOS
near the threshold was obtained in \cite{voit,wiegmann}.
Parenthetically, functional form (\ref{mottdos}) remains valid when 
the spin channel is gapped 
as well.
In this case the density of states of spin excitations is given by 
Eq.(\ref{massivedos})
with $\Delta \rightarrow \Delta_{\sigma}$. Peforming integration in 
(\ref{convolution}),
we again find behavior described by Eq.~(\ref{mottdos}) near the threshold
$\Delta +  \Delta_{\sigma}$.
In other words, the gap in the electron's DOS is given by the sum of charge-
and spin-sector gaps.

\subsection{DOS of the electron ladder.}
\label{electronladder}

We consider now the tunneling density of states of a two-channel wire.
To apply Eq.(\ref{gapped-corr}), one needs to have the strong-coupling
fixed-point values of $K_{\nu}$, which are not known for a general case of
two non-equivalent channels coupled by the Coulomb interaction. Just to 
illustrate
what kind of behavior one might expect in this case, we
consider an electron ladder with $
K_{1\nu}=K_{2\nu} ~(\nu=\rho,\sigma)$ in the Cooper phase.
The right-moving fermion is represented by
\begin{equation}
R_{n=1,s}=\frac{e^{ik_F x}}{\sqrt{2\pi\alpha}} ~e^{is\sqrt{\pi}
(\varphi_{\sigma-} -\theta_{\sigma-})/2} ~e^{is\sqrt{\pi}(\varphi_{\sigma+}
- \theta_{\sigma+})/2} ~e^{i\sqrt{\pi}(\varphi_{\rho-}-\theta_{\rho-})/2}
~e^{i\sqrt{\pi}(\varphi_{\rho+}-\theta_{\rho+})/2}.
\end{equation}
Now we apply Eq.(\ref{gapped-corr}) to the correlator 
$F(\tau)=-\langle T_{\tau}R_{1,s}(\tau,0)
R^{\dagger}_{1,s}(0,0) \rangle$. In the Cooper phase,
$\theta_{\rho-}$ and $\varphi_{\sigma \pm}$ are
gapped, hence their conjugates are exponentially suppressed. 
As a consequence, e.g.,
\begin{equation}
\langle e^{i\sqrt{\pi}(\varphi_{\sigma-}(\tau) - \theta_{\sigma-}(\tau))/2} 
~e^{-i \sqrt{\pi}(\varphi_{\sigma-}(0) - \theta_{\sigma-}(0))/2}\rangle
=C_{\sigma}\Big( \frac{\alpha}{v_{\sigma}|\tau|}\Big)^{1/8} 
\exp\left[-\frac{\pi m_{\sigma}v_{\sigma}|\tau|}{16}\right].
\end{equation}
On the other hand,
$\varphi_{\rho+}$ and $\theta_{\rho+}$  remain critical. As a result, 
\begin{equation}
F(\tau)\propto sgn(\tau)~ |\tau|^{-\kappa}
 \exp\left[-\frac{\pi}{16}(2 m_{\sigma}v_{\sigma} + 
m_{\rho-} v_{\rho-})|\tau|\right], 
~\kappa=\frac{1}{8}(3 + K_{\rho+} + 1/K_{\rho+}).
\end{equation}
 Hence the DOS behaves as
\begin{equation}
\rho(\omega) \propto \frac{\Theta(\omega - \Delta_{SC})}
{(\omega - \Delta_{SC})^{\gamma}},
~\gamma=1 - \kappa=\frac{1}{8}(5 - K_{\rho+} - 1/K_{\rho+}),
\label{ladder_dos}
\end{equation}
where $\Delta_{SC}=(\pi/16)(2m_{\sigma} v_{\sigma} + m_{\rho-} v_{\rho-})$.
Note that $\gamma \leq 3/8 < 1/2$ for $K_{\rho+} \leq 1$.

Exactly at half-filling, when $\rho+$ mode is also gapped due to Umklapp
scattering and there are no more gapless modes, our procedure gives $F(\tau) 
\sim e^{-\Delta|\tau|}/\sqrt{|\tau|}$,
in agreement with recent exact result \cite{konik}. The
corresponding DOS is that of a free massive particle, $\rho_{hf}(\omega)
\sim \Theta(\omega - \Delta)/\sqrt{\omega - \Delta}$.

Comparison of (\ref{ladder_dos}) with (\ref{mottdos}) shows that softening
of the square-root singularity is pronounced weaker for a ladder than
than for  a Hubbard chain, because a ladder has three
gapped and only one gapless mode.
 As repulsion in the $\rho+$ channel becomes 
stronger,
i.e., as $K_{\rho+}$ decreases, the singularity becomes weaker and disappears
at $K_{\rho+}=\left(5-\sqrt{21}\right)/2\approx 0.2$. For even smaller 
$K_{\rho+}$,
we have $\rho(\omega=\Delta_{SC})=0$. This behavior though is not very 
realistic as it requires
very strong repulsion. For weak repulsion, i.e., when $K_{\rho+}\approx 1$,
\begin{equation}
\rho(\omega)\propto \Theta(\omega-\Delta_{SC})/(\omega-\Delta_{SC})^{3/8}, 
\end{equation}
and the threshold
singularity is still present albeit softened compared to the free-massive-particle case. We note that $\rho(\omega)$ (\ref{ladder_dos}) is
similar to the DOS of high unoccupied subbands of a quantum wire,
considered recently by Balents \cite{balents3}.

One would expect that the long-range order of the Cooper phase
affect tunneling.
Indeed one finds that {\it pair correlations} are determined
by the Luttinger parameter of the total charge fluctuations only
\cite{balents},
\begin{equation}
\langle R_{1,s}(\tau)L_{1,-s}(\tau)~L_{1,-s}^{\dagger}(0)
R_{1,s}^{\dagger}(0)\rangle
\sim \langle e^{-i\sqrt{\pi}\theta_{\rho+}(\tau)}e^{i\sqrt{\pi}
\theta_{\rho+}(0)}\rangle
\sim \tau^{-1/(2K_{\rho+})},
\end{equation}
whereas all other two-particle combinations decay exponentially. Thus,
although the single-particle density of states is strictly zero, 
the two-particle
one is not. In principle, this effect can be checked experimentally by
tunneling into a two-channel wire from the superconducting tip - one should 
observe then a nonzero tunneling current of Cooper pairs.
Its magnitude, however, will be much smaller than  the current in a system  of
a normal tip and gapless wire, because the probability of two-particle
(Copper pair) tunneling ($|{\cal T}|^4$) is exponentially smaller than that
of single-particle tunneling ($|{\cal T}|^2$).

\section{Tunneling into the end of a gapped wire}
\label{end-LL}

Tunneling into the end of a Luttinger liquid is different
from tunneling into the bulk. The reason
for this difference is the open boundary condition $\psi =0$ for the electron
wavefunction. For boson modes describing charge
and spin displacements, this condition means pinning at the boundary.
The difference between the edge and bulk tunneling was considered first
theoretically by Kane and Fisher \cite{kane_fisher}, and has recently
been observed in experiments on tuneling into carbon nanotubes \cite{cobden}.
A rigorous treatment of a Luttinger liquid with
open boundary conditions, which involves re-formulation
of the bosonization procedure, can be found in Refs.~\cite{fg,fradkin,gnt}.

Suppose now that a two-subband wire is driven into a CDW state by
direct backscattering processes accompanied by density adjustment, as
described in Sec.~\ref{cdw-case}. The relative mode of charge
excitations is described by Hamiltonian (\ref{H-}), in which we put 
$\delta k_F=0$. From the equivalence of Eqs.~(\ref{umklapp}) and (\ref{H-}) 
(with $\delta k_F=0$),
we expect the fixed point-value of $K_{-}$ in the CDW-phase to be the
same as for a half-filled Hubbard chain, i.e., $K_{-}^{*}=1/2$. The total charge
mode ($\varphi_+$) remains gapless and plays the same role as the spin mode of the
Hubbard chain [see Eqs.~(\ref{mottdos}) and (\ref{ladder_dos})]:
it softens the threshold singularity of the DOS. For $\omega<\Delta_{CDW}$, 
the density of states is equal to zero.
Thus if the tunneling contact probes the interior of the wire, a
gapped behavior is observed.

However, the DOS at the end of a wire exhibits {\it gapless} behavior, i.e, 
$\rho_{end}\propto
|\omega|^{\alpha_d}$, as we will demonstrate in the rest of this Section.
Consequently, $I(V)\propto |V|^{\alpha_d+1}$ for tunneling from a macroscopic 
(Fermi-liquid contact)
into the end of the wire, and $I(V)\propto|V|^{2\alpha_d+1}$ for tunneling
through a barrier located somewhere within the wire.

This very different behavior of the DOS in the bulk and at the
end of the wire can be understood physically for tunneling through
a barrier located within the wire (cf. Fig.~\ref{fig:cdw}).
Without the barrier, the CDW is
free to slide and the conductance is the same as in the
absence of any interactions. Squeezing one
electron into the middle of the wire leads
to creation of a soliton-like compression in one of the two modes, and to
accompanying it ``stretch'' in  the other mode, which
requires an energy of the order of the charge gap
$\Delta_{CDW}$. However, such an excitation needs not be created when 
the barrier distorts the uniform profile of the CDW.
 Indeed, the boundary condition imposed by the
barrier pins the mode $\varphi_{-}$ at $x=0$ to a value which is
different from the one in the bulk, $\varphi_{-}(x\to\infty)=\sqrt{\pi/8}$ (the
latter follows from the  minimization  of the CDW energy, as
illustrated in Fig.\ref{fig:cdwbarrier}). Therefore, the regular order
of CDW is already frustrated near the barrier: there is a built-in
compression in one of the modes, and depression in the other. The
electron that tunnels through the barrier arrives into the
\lq\lq stretched\rq\rq\/ mode. Upon the proper shift of both modes, the system
arrives into a state with the same energy but with a switched
\lq\lq polarity\rq\rq\/ of frustration. This consideration is true if the barrier
is strong enough to 
destroy the CDW order in its vicinity, i.e., the barrier height is larger
than $\Delta_{CDW}$. 


\begin{figure}[t]
\hspace*{5cm}
\epsfxsize=5cm
\epsfbox{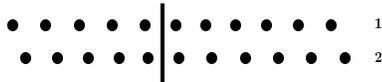}
\vskip0.5cm
\caption{Illustration of the CDW in the vicinity of a
barrier~: while the bulk value for $\varphi_{-}$ is
$\sqrt{\pi/8}$, the barrier forces $\varphi_{-}$
to vanish at $x=0$.}
\label{fig:cdwbarrier}
\end{figure}

We now give a derivation for the current through the barrier.
To model the boundary conditions corresponding to a wire cut
into two semi-infinite pieces, we choose the potential barrier 
in the form $w(x,\rperp)=\pi\alpha W \delta(x)$, so that
$W_{n=1,2}=\pi\alpha W$ and $W_{inter}=0$
[cf. Eqs.~(\ref{wndef},\ref{winterdef})]. 
To find the current in, e.g., the 1st subband, through the barrier at $x=0$
\beq\label{rate}
I\propto\lim_{t\to\infty}\la\partial_t\varphi_1(x=0,t)\ra/e,
\eeq
we need to calculate the rate at which  $\varphi_1(x=0,t)$
increases in the stationary limit in the presence of a potential 
drop $\sim -eV\int dx\;
(\partial_x\varphi_{+}/2\sqrt{\pi}){\rm sgn}(x)$, proportional
to the applied voltage $V$ (due to their equivalence, it
does not matter which of the two modes we use for measuring the
current).
Eq.\ (\ref{rate}) reduces the transport problem to that  of the dynamics
of a quantum particle $q(t)\equiv\varphi_1(x=0,t)$
subject to \lq\lq damping\rq\rq\/ by all of the remaining bulk degrees of
freedom, including those of the second mode. Therefore, we can employ
methods of dissipative quantum mechanics
\cite{weiss,kane_fisher} to solve this problem.

The effective action of the boundary mode $S_{eff}$ is obtained by
tracing over these remaining degrees of freedom
\beq\label{trace}
e^{-S_{\rm eff}[q]}=\int{\cal D}[\varphi_1]{\cal D}[\varphi_2]\;
\delta[q(\tau)-\varphi_1(x=0,\tau)]\;\delta[\varphi_2(x=0,\tau)]\;e^{-S},
\label{action-eff}
\eeq
where
\bea
S=\int dx\int d\tau&&\Bigl[\frac{1}{2v_F}(\partial_{\tau}\varphi_{+})^2
+\frac{v_F}{2K_{+}^2}(\partial_{x}\varphi_{+})^2\\[3ex]
&&\mbox{}+\frac{1}{2v_F}(\partial_{\tau}\varphi_{-})^2
+\frac{v_F}{2K_{-}^2}(\partial_{x}\varphi_{-})^2
+\frac{4f_{bs}}{\pi\alpha^2}\varphi_{-}^2\Bigr]\;,
\eea
where $\pm$ combinations are defined in (\ref{pm}).
Performing the integration, we get
\beq\label{seff}
S_{\rm eff}[q]=\frac{1}{2}\int d\tau\int d\tau'\;q(\tau)
\hat{{\cal K}}(\tau-\tau')q(\tau'),
\eeq
which accounts for the influence of the bulk modes exactly. In
(\ref{trace}) we assume a large barrier so that incoherent
single electron passages through subband 1 determine the current,
while subband 2 is fixed by the barrier, cf.\ (\ref{action-eff}).
Simultaneous contributions from subband 2 represent coherence effects, and are
of higher order in the barrier transmission coefficient. The dynamics of
$q$ is governed by
\beq\label{s0}
S_0=\int d\tau\;\Bigl(W\cos 2\sqrt{\pi}q
-\frac{eV}{\sqrt{2\pi}}q\Bigr)+S_{\rm eff},
\eeq
which includes the applied voltage. Eq.\ (\ref{s0}) describes a
damped quantum particle on a tilted washboard potential. Tunneling
between adjacent minima of the potential corresponds to single 
electron transfers.

For large $W$, Eq.\ (\ref{s0}) maps onto a tight binding model
with nearest neighbor hopping ${\cal T}$, whose value can
in principle be deduced from $W$, assuming a $\delta$-barrier~:
it is renormalized by damping compared to the bare value
\cite{weiss96}. In this analogy, (\ref{rate}) corresponds to
the particle's mobility, studied in Ref.\cite{weiss91}. In
leading order $\sim|{\cal T}|^2$, the result is
\beq
I(V)=e|{\cal T}|^2\int_0^{\infty}dt\;\sin(eVt)\:{\rm Im}\:e^{-w(t)},
\eeq
where
\beq
w(t)=\int_0^{\infty}\frac{J(\omega)}{\omega^2}(1-e^{-i\omega t})\;.
\eeq
The spectral function $J(\omega)$ is related to the Fourier
transform ${\cal K}(\omega_n)$ of the kernel \cite{weiss}
appearing in (\ref{seff}),
\beq
J(\omega)=-\lim_{\eta\to 0}{\rm Im}\:{\cal K}(-i\omega+\eta)=
\frac{\omega}{K_{+}}+
\frac{\sqrt{\omega^2-\omega_0^2}}{K_{-}}\:\Theta(\omega-\omega_0),
\eeq
with $\omega_0^2=8f_{bs}v_F/\pi\alpha^2$. This yields
\beq
I(V)\sim V^{1/K_{+}-1},
\label{i-v}
\eeq
which gives $\alpha_d=1/2K_{+}-1$ at voltages $V<\omega_0/e$. Above the gap, 
i.e, 
for $V\gg\omega_0/e$, gapless behavior of a two-subband Luttinger liquid
is restored $\:I(V\gg\omega_0/e)\sim V^{1/K_{+}+1/K_{-}-1}\:$.

The  gapless behavior of fermion's DOS
at the end of the wire can be interpreted in terms of a
 midgap state in the
$(-)$ channel localized near the boundary. This is a bound
state with zero energy formed in the potential well created by 
the static distortion of the $\varphi_{-}$
field subject to an open boundary condition $\varphi_{-}(x=0)=0$ (see
 Appendix D of  the paper by Fabrizio and Gogolin \cite{fg}).
Qualitatively, this state is similar to the boundary state in a doped two-leg
spin ladder, which represents free $S=1/2$ spin induced by a charged impurity,
 see, e.g., \cite{doped_ladder}.
As a result, {\it local} density of states of the $(-)$ channel 
at the end of the wire takes the form 
$\rho_{-}(\epsilon)\sim \lambda \delta(\epsilon) + 
\rho_-^{\text{reg}}$, where $\rho_-^{\text{reg}}$ stands for contribution 
of massive modes with energies {\it above} the CDW gap.
On the other hand, the $(+)$ mode remains gapless, and its end-chain
DOS is given by
$\rho_{+}(x=0,\omega) \sim \omega^{\frac{1}{2K_{+}}-1}$,
~\cite{kane_fisher,fg,gnt}.
The factor of $1/2$ in the exponent is due to \lq\lq
factorization\rq\rq
 of the electron operator into $(\pm)$ modes. Another consequence of such
factorization is that the DOS of a physical electron
is a convolution of the DOS of the $(\pm)$ channels
[cf. Eq.~(\ref{convolution})]
\beq
\rho_{phys}(x=0,\omega) \sim \int_0^{\omega} d\epsilon \rho_{+}(x=0,\epsilon)
\rho_{-}(x=0,\omega-\epsilon).
\eeq
Therefore,
\beq
\rho_{phys}(x=0,\omega) \sim \rho_{+}(x=0,\omega)=
\omega^{\frac{1} {2K_{+}} - 1},
\eeq
which implies  $\alpha_{\rm d}=1/(2K_{+})-1$, in agreement 
with the result of the explicit calculation of $I(V)$ presented above,
Eq.~(\ref{i-v}).
\section{Experimental consequencies and Conclusions}

Apart from the rather well-known Luttinger-liquid phase, a two-subband
quantum wire may also exhibit either a CDW or  
superconducting (Cooper) phase.
In both of these phases, certain modes of inter-subband charge- and
spin-excitations are gapped, whereas the center-of-mass charge mode
remains gapless. As a result, the conductance remains at the universal value
of $2e^2/h$ per occupied subband, irrespective of whether the wire
is in a gapless or gapped phase.
However, the single-particle density of states in the middle of the wire
has a hard gap. Above the gap, the DOS exhibits a non-universal
threshold behavior
$\rho(\omega)\sim \Theta(\omega - \Delta)(\omega - \Delta)^{-b}$,
where $b
\leq 1/2$. Softening of the threshold singularity is due to
 \lq\lq dressing\rq\rq\/ of 
gapped modes by the remaining gapless one.

We find that the DOS for tunneling into the end of a wire in the CDW phase 
remains gapless, with the exponent determined by the center-of-mass mode only.
This effect is due to frustration introduced into the CDW order
by an open boundary (strong barrier).

Where should one look for such exotic phases of a quantum wire?
We believe that quantum wires \cite{yacoby} prepared by cleaved edge
overgrowth technique may be well-suited  for observing the CDW phase. 
Indeed, the cross-section of such a wire is close to 
a square,  which implies that the lowest states of transverse
quantization should be close in energy.
Hence, one-dimensional subbands can have close
Fermi-momenta, which is a necessary condition for the formation
of the CDW state. The Cooper phase, on the other hand, requires
effective attraction in the relative charge channel,
and has the best chance to occur  when the second (upper) subband
just opens for conduction. i.e., is near the Van Hove singularity.

\acknowledgements
We would like to thank A. Abanov, B. Altshuler,  V. J. Emery, 
M. Reizer, B. Marston, 
A. Nersesyan, T. M. Rice, S. Sachdev, and N. Shannon for interesting
discussions and G. Martin for his help in manuscript
preparation.  DLM acknowledges the financial support from NSF DMR-970338.
WH acknowledges kind hospitality of the University of Minnesota
and support from the
DFG (Germany) through contract HA 2108/4-1.
The work at the University of Minnesota was supported by NSF Grants
DMR-9731756 and DMR-9812340.
DLM and LIG would like to thank the organizers of the workshop at Centro Stefano Franscini, 
Switzerland, 
where part of this work has been done, and the organizers of the WHE workshop
in Hamburg.

\references
\bi{schulz_lezush} H. J. Schulz, in {\it Proceedings of Les Houches Summer School LXI}, 
ed. E. Akkermans, G. Montambaux, J. Pichard, and J. Zinn-Justin 
(Elsevier, Amsterdam, 1995), p.533.
\bibitem{fisher_glazman} M. P. A. Fisher and L. I. Glazman in {\it{Mesoscopic
Electron Transport}}, edited by L. P. Kouwenhoven, L. L. Sohn and G. Sch\"on,
Kluwer Academic, Boston, 1997.
\bibitem{cobden} M. Bockrath et al., Nature {\bf 397}, 598 (1999); 
cond-mat/9812233.
\bibitem{dekker} C. Dekker, Physics Today {\bf 52}, 22 (1999).
\bibitem{yacoby_new} C. Ilani et al., cond-mat/9910116.
\bibitem{zaitsev} S.\ V.\ Zaitsev-Zotov et al., cond-mat/990756.
\bibitem{kane_fisher} C. L. Kane and M. P. A. Fisher, \prb {\bf 46}, 15233 (1992).
\bibitem{matveev_glazman} K. A. Matveev and L. I. Glazman, \prl {\bf 70}, 990
(1993).
\bi{tarucha} S.\ Tarucha, T.\ Honda, and T. Saku, Sol. St. Commun. {\bf 94},
 413 (1995).
\bi{pepper} K. J. Thomas et al., \prl {\bf 77}, 135 (1996).
\bi{yacoby} A. Yacoby et al., \prl {\bf 77}, 4612 (1996); Solid State
Communications {\bf 101}, 77 (1997).
\bi{finkelstein} A.\ M.\ Finkelstein and A.\ I.\ Larkin, \prb {\bf 47}, 10461 (1993).
\bi{fabrizio} M.\ Fabrizio, \prb {\bf 48}, 15838 (1993).
\bibitem{schulz2} H. J. Schulz, \prb {\bf 53}, R2959 (1996).
\bibitem{leon2} L. Balents and M. P. A. Fisher, \prb {\bf 53}, 12133 (1996).
\bibitem{schulz} H. J. Schulz, cond-mat/9808167.
\bibitem{ekz} V. J. Emery, S. A. Kivelson, and O. Zachar,
\prb {\bf 56}, 6120 (1997).
\bibitem{balents} H. H. Lin, L. Balents, and M. P. A. Fisher, \prb {\bf 56},
6569 (1997); cond-mat/9801285.
\bi{kane} C. L. Kane, L. Balents and M. P. A. Fisher, \prl {\bf 79}, 5086
(1997).
\bi{krotov} Yu.\ A.\ Krotov, D.\ H.\ Lee. and S.\ G.\ Louie, \prl {\bf 78},
4245 (1997).
\bi{egger} R.\ Egger and A.\ O.\ Gogolin, \prl {\bf 79}, 5082 (1997).
\bibitem{gogolin} R. Egger, A. O. Gogolin, Eur. Phys. J. B {\bf 3}, 281
(1998).
\bi{odintsov} H.\ Yoshioka and A.\ Odintsov, \prl {\bf 82}, 374 (1999). 
\bi{prigodin} Yu.\ A.\ Firsov, V.\ N.\ Prigodin, and Chr. Seidel, Phys. Rep.
{\bf 126}, 245 (1985). 
\bibitem{varma} C. M. Varma and A. Zawadowski, \prb {\bf 32}, 7399 (1985).
\bibitem{voit} J. Voit, Eur.\ Phys. J.\ B {\bf 5}, 505 (1999).
\bibitem{wiegmann} P. B. Wiegmann, \prb {\bf 59}, 15705 (1999).
\bibitem{orignac} E. Orignac and T. Giamarchi, \prb {\bf 56}, 7167 (1997).
\bibitem{mori} M. Mori, M. Ogata and H. Fukuyama, J. Phys. Soc. Jpn.
{\bf 66}, 3363 (1997).
\bibitem{starykh_maslov} O. A. Starykh and D. L. Maslov, \prl {\bf 80}, 1694 
(1998).
\bibitem{kouwenhoven} L. P. Kouwenhoven et al., \prl {\bf 65}, 361 (1990).
\bi{efetov} K.\ B.\ Efetov and A.\ I.\ Larkin, Zh.\ Exp.\ Teor.\ Phys.\ 
{\bf 69}, 764 (1975) [Sov.\ Phys.\ -JETP {\bf 42}, 390 (1976)].
\bibitem{froelich} H.\ Fr\"olich, J. Phys. C {\bf 1}, 544 (1968).
\bibitem{ruvalds} J.\ Ruvalds, Adv.\ Phys. {\bf 30}, 677 (1981).
\bi{haldane} F.\ D.\ M.\ Haldane, \prl {\bf 45}, 1358 (1980).
\bi{giam2} T. Giamarchi, \prb {\bf 44}, 2905 (1991).
\bi{metzner} W.\ Metzner and C.\ Di Castro, \prb {\bf 47}, 16107 (1993).
\bi{mahan} G. D. Mahan, {\it Many-Particle Physics}, 2nd ed., Plenum Press,
New York, 1990, section 4.4.
\bibitem{luttinger} J. M. Luttinger, J. Math.. Phys. {\bf 4}, 1154 
(1963).
\bi{emery_review} V. J. Emery, {\it{Highly Conducting One-Dimensional Solids}},
eds. J. T. Devreese, R. E. Evrard and V. E. van Doren, New York, Plenum,
p. 247 (1979).
\bibitem{capponi} S. Capponi, D. Poilblanc and T. Giamarchi, cond-mat/9909360.
\bibitem{matveev93} K.A. Matveev, D. Yue, and L.I. Glazman, \prl {\bf 71},
3351 (1993).
\bibitem{pokr} V. L. Pokrovsky and A. L. Talapov, Sov. Phys. JETP 
{\bf 48}, 570 (1978).
\bibitem{c-ic} H. J. Schulz, \prb {\bf 22}, 5274 (1980).
\bibitem{luther} A. Luther, \prb {\bf 15}, 403 (1977).
\bibitem{giam} T. Giamarchi and H. Schulz, \prb {\bf 39}, 4620 (1989)
\bibitem{kogut} J. B. Kogut, Rev. Mod. Phys. {\bf 51}, 701 (1979).
\bibitem{safi} I. Safi and H. J. Schulz, \prb {\bf 52}, R17040 (1995).
\bibitem{maslov-stone} D. L. Maslov and M. Stone, \prb {\bf 52}, R5539
(1995).
\bibitem{ponomarenko} V. V. Ponomarenko, \prb {\bf 52}, R8666 (1995).
\bibitem{giam_book} T. Giamarchi and H. Maurey, in {\it Correlated
Fermions and Transport in Mesoscopic Systems}, edited by T. Martin,
 G. Montambaux, and J. Tran Thanh Van (Editions Frontieres, 1996), p. 13.
\bibitem{dzyal} I. E. Dzyaloshinskii and A. I. Larkin, Sov. Phys. JETP
{\bf 38}, 202 (1974).
\bi{chang} A. M. Chang, L. N. Pfeiffer, and K. W. West, \prl {\bf 77}, 
2538 (1996).
\bi{shytov} A. V. Shytov, L. S. Levitov, and B. I. Halperin, \prl {\bf 80}, 141
 (1998).
\bi{alekseev}A. Alekseev, V. Cheianov, A. P. Dmitriev,
  and V. Yu. Kachorovskii, cond-mat/9904076.
\bibitem{chui} S. T. Chui and P. A. Lee, \prl {\bf 35}, 315 (1975).
\bibitem{gulacsi} Z. Gulacsi and K. S. Bedell, \prl {\bf 72}, 2765 (1994).
\bibitem{luther-emery} A. Luther and V. J. Emery, \prl {\bf 33}, 589 (1974).
\bi{sachdev} S. Sachdev, T. Senthil and R. Shankar, \prb {\bf 50}, 258 (1994).
\bibitem{shelton} D. G. Shelton, A. A. Nersesyan, and A. M. Tsvelik, \prb
{\bf 53}, 8561 (1996).
\bibitem{tsvelik2} A. M. Tsvelik, {\it{Quantum Field Theory in
Condensed Matter Physics}}, Cambridge University Press, 1995.

\bibitem{konik} R. Konik, F. Lesage, A. W. W. Ludwig, H. Saleur,
cond-mat/9806334.
\bibitem{balents3} L. Balents, cond-mat/9902159.
\bibitem{fg} M. Fabrizio and A. O. Gogolin, \prb {\bf 51}, 17827 (1995).
\bi{fradkin} M. Fuentes, A. Lopez, E. Fradkin, and E. Moreno, Nucl. Phys. B
{\bf 450}, 603 (1995).
\bibitem{gnt} A. O. Gogolin, A. A. Nersesyan and A. M. Tsvelik, 
{\it{Bosonization and Strongly Correlated Systems}}, Cambridge University
Press, 1998.
\bibitem{weiss} U. Weiss, {\em Quantum Dissipative Systems},
Vol.~2 of {\em Modern Condensed Matter Physics}, World
Scientific, Singapore, 1993; M. Sassetti and U. Weiss,
Europhys.~Lett. {\bf 27}, 311 (1994).
\bibitem{weiss96} U. Weiss, Solid State Comm. {\bf 100}, 281 (1996).
\bibitem{weiss91} U. Weiss et al., Z. Phys. B {\bf 84}, 471 (1991).
\bibitem{doped_ladder} A. O. Gogolin, A. A. Nersesyan, A. M. Tsvelik, and
Lu Yu, Nucl. Phys. B {\bf 540}, 705 (1999).

\end{document}